%% file: plos_latex_template_AF.tex
\documentclass[10pt,letterpaper]{article}
\usepackage[top=0.85in,left=2.75in,footskip=0.75in]{geometry}

\usepackage{amsmath,amssymb}

\usepackage{changepage}

\usepackage[utf8x]{inputenc}

\usepackage{textcomp,marvosym}

\usepackage{cite}

\usepackage{nameref}
\usepackage[colorlinks=true,filecolor=blue,urlcolor=blue,citecolor=blue,bookmarksnumbered]{hyperref} 
\usepackage[right]{lineno}

\usepackage{microtype}
\DisableLigatures[f]{encoding = *, family = * }

\usepackage[table]{xcolor}

\usepackage{array}

\newcolumntype{+}{!{\vrule width 2pt}}

\newlength\savedwidth

\newcommand\thickhline{\noalign{\global\savedwidth\arrayrulewidth\global\arrayrulewidth 2pt}%
\hline
\noalign{\global\arrayrulewidth\savedwidth}}

\raggedright
\usepackage[aboveskip=1pt,labelfont=bf,labelsep=period,justification=raggedright,singlelinecheck=off]{caption}
\renewcommand{\figurename}{Fig}

\makeatletter
\renewcommand{\@biblabel}[1]{\quad#1.}
\makeatother

\usepackage{lastpage,fancyhdr,graphicx}
\usepackage{epstopdf}
\fancyheadoffset[L]{2.25in}
\fancyfootoffset[L]{2.25in}
\usepackage[normalem]{ulem}
\usepackage{soul} 
\setul{}{0.25ex} 

\newcommand{\todayd}{%
\the\year/{\ifnum \month < 10 0\the\month \else \the\month \fi}/%
{\ifnum \day < 10 0\the\day \else \the\day \fi}}
\newcommand{\todaye}{\the\year--\the\month--\the\day}

\newcommand{\Err}{\text{Err}} 

\usepackage{ifthen}
\newcounter{verbosity}
\definecolor{Lightergray}{rgb}{0.75,0.75,0.75}
\definecolor{DarkerGreen}{rgb}{0,0.6,0}
\colorlet{StrikeoutColor}{blue!50}

\newcommand{\ModifiedRF}[1]{\textcolor{blue}{#1}}
\newcommand{\AddedRF}[1]{\ModifiedRF{#1}}

\newcommand{\ModifiedKA}[1]{\textcolor{DarkerGreen}{#1}}
\newcommand{\AddedKA}[1]{\ModifiedKA{#1}}
\newcommand{\NotesKA}[1]{\ModifiedKA{[@@@@ KA: #1 @@@@]}}

\newcommand{\LDeletedKA}[1]{\textcolor{Lightergray}{#1}}

\newcommand{\hlt}[1]{\textcolor{black}{#1}}
\usepackage{upgreek} 

\let\cite\cite
\let\citep\cite

\graphicspath{%
  {./}}
  
\usepackage{placeins}
\begin{document}
\vspace*{0.2in}

\begin{flushleft}
{\Large
\textbf\newline{A model for the size distribution of marine microplastics: a statistical mechanics approach} 
}
\newline
\\
Kunihiro Aoki\textsuperscript{1\Yinyang},
Ryo Furue\textsuperscript{1\Yinyang},
\\
\bigskip
\textbf{1} Japan Agency for Marine-Earth Science and Technology, Yokohama, Kanagawa, Japan
\\
\bigskip

%
%
\Yinyang These authors contributed equally to this work.





* kaoki@jamstec.go.jp

\end{flushleft}
\section*{Abstract}
The size distribution of marine \hlt{microplastics} provides a fundamental data source for understanding the dispersal, break down, and biotic impacts of the microplastics in the ocean. The observed size distribution \hlt{at the sea surface} generally shows, from large to small sizes, a gradual increase followed by a rapid decrease. This decrease has led to the hypothesis that the smallest fragments are selectively removed by sinking or biological uptake. Here we propose a new model of size distribution without any removal of material from the system.
The model uses an analogy with
black-body radiation
and the resultant size distribution is analogous to Planck's law. In this model, the original large plastic piece is broken into smaller pieces once by the application of ``energy'' or work by waves or other processes, under two assumptions, one that fragmentation into smaller pieces requires larger energy and the other that the probability distribution of the ``energy'' follows the Boltzmann distribution. Our formula well reproduces observed size distributions over wide size ranges from micro- to \hlt{mesoplastics}. According to this model, the smallest fragments are fewer because large ``energy'' required to produce
such small fragments occurs more rarely.



\section*{Introduction}\label{sec:intro}

A large fraction of the estimated billion tonnes of plastic waste that goes into the ocean\citep{JambeckEA2015plastic} is found in a fragmented form, ``microplastics'', with sizes of less than 5\,mm \citep{BarnesEA2009plastic} through photodegradation and weathering processes\citep{CorcoranEA2009plastic, Andrady2011plastic, Andrady2017}. Those microplastics spread globally\citep{ EriksenEA2014plastic, SebilleEA2015plastic, IsobeEA2019plastic, OninkEA2019plastic},
potentially acting as a transport vector of chemical pollutants\citep{AshtonEA2010plastic, HolmesEA2012plastic, NakashimaEA2012plastic} and causing physical and chemical damages on marine organism\citep{BrowneEA2008, BoergerEA2010plastic, MurrayEA2011plastic}. Recent drift simulations of microplastics calibrated against observed abundance of microplastics have produced global or semi-global maps of estimated microplastics abundance and concentration
near the sea surface\citep{EriksenEA2014plastic, IsobeEA2019plastic, SebilleEA2015plastic}.
These simulations will be further used to assess the biological impacts
of microplastics. 

Such simulations generally assume the size distribution of microplastics
and their results would depend on the assumption because sedimentation and biological uptake can depend on size\citep{BoergerEA2010plastic,JabeenEA2017plastic, IwasakiEA2017plastic, SagawaEA2018plastic}.
Interestingly, the number of small pieces rapidly decreases toward smaller sizes \hlt{of $O$(100\,$\upmu$m)} in most observations \hlt{of plastic fragments at the sea surface}\citep{IsobeEA2014plastic, CozarEA2014plastic, IsobeEA2015plastic, EoEA2018plastic}.
This feature is puzzling
because the number of plastic pieces is expected to increase toward smaller sizes
if the pieces keep broken down into smaller and smaller pieces (progressive fragmentation).
For example, C{\'o}zar et al.\cite{CozarEA2014plastic} indicates that a type of progressive fragmentation
leads to a cube law toward smaller sizes.
The observed decrease at smaller sizes
has generally been hypothesized to be due to selective sinking to depths, \hlt{sampling error}, or
to selective ingestion by marine organisms\citep{CozarEA2014plastic, IsobeEA2014plastic, IsobeEA2015plastic, ReisserEA2015plastic,EndersEA2015plastic,KooiEA2019plastics}.

The fracture mechanisms of marine plastics, however, are not well known.
The power law\hlt{, which results from scale-invariant fracture processes,} is often invoked to explain observed size
distributions of plastics.
It tends to fit well observed size distributions
at larger sizes\cite{CozarEA2014plastic}.
The power law can be derived, for example, from
collision cascade among objects,
as often applied to the fragmentation of asteroids\citep{Dohnanyi1969, DavisEA1989, TanakaEA1996},
which does not include any decrease toward smaller sizes unless or until other processes than fragmentation start to dominate. Somewhat different fragmentation processes lead to a log-normal distribution\citep{Kolmogorov1941lognormal, Middleton1970, Crow+Shimizu1987lognormal},
which \hlt{is premised on the size reduction rate following the Normal distribution and} has been applied to the fragmentation of sand grains,
mastication,
and the fracturing of thin glass rods\citep{Vincent1986, KobayashiEA2006mastication, IshiiEA1992},
but not to oceanic microplastics to our knowledge.
The log-normal distribution has a peak skewed toward smaller sizes and 
is similar to observed size distributions of fragments at smaller sizes
but tends to deviate from observed distribution at larger sizes, where the power law
tends to fit better\cite{CozarEA2014plastic}.
To the best of our knowledge,
only the Weibull distribution\citep{Weibull1951, BrownEA1995} is similar to observed size distributions across the microplastics and mesoplastics ranges \hlt{(fragments larger than 5\,mm)},
at least in C\'{o}zar et al.'s study\cite{CozarEA2014plastic}.
The Weibull distribution as applied to the size distributions
of various types of particles is largely empirical but with an interpretation as resulting
from the branching tree of cracks\cite{BrownEA1995}.

In this paper, we propose a new model where ``energy'' needed to break down the plastic pieces acts as a constraint. We assume that the probability of the ``energy'' obeys the Boltzmann distribution. The statistics of fragmentation is then analogous to that of the black body radiation and the resultant size distribution is analogous to the Planck distribution. As we shall see, \hlt{without the assumption of selective sinking or ingestion}, this size distribution has a peak skewed toward small sizes and the smaller plastic pieces beyond the peak are much scarcer. \hlt{We also discuss the generation of finer microplastics ($O$(10\,$\upmu$m)), which is recently observed mostly in the mixed layer, in terms of our fracture model.}

\section*{Fracture model}
\hlt{We assume that plastic waste first becomes fragile on beaches
because of ultraviolet light and other weathering factors
and then is broken up on beaches into microplastic pieces by the action of waves, winds, and other forces
before being washed off into the ocean.
Accordingly, the amount (mass) of microplastics which is produced on beaches
will be determined by these processes.
The fraction of microplastics that goes into the ocean may depend on waves and other weather factors.
The microplastics will then be diluted in the ocean
by mixing due to turbulence.
These are the microplastics which observations sample.
In the following we build a simple, idealized statistical model that determines the size distribution of the microplastics produced on the beaches.
The model considers the fragmentation of given plastic mass and is naturally mass-conserving.}

We offer the following simple physical model only
as a representation of complicated fragmentation processes such as one-time crush and slow weathering. 
We do not claim that the following are exactly
the underlying processes that lead to the observed size distribution.
\hlt{This model is originally intended to explain microplastics larger than $O$(100\,$\upmu$\,m). It will be later extended for finer microplastics (see the Discussion section and \ref{sec:3d_model}).
In this section, we only show an outline of the derivation. Details are found in ``\nameref{sec:method_theory}'' of the Materials and Methods section.}

\begin{figure*}
  \centering\includegraphics[width=1.0\textwidth]{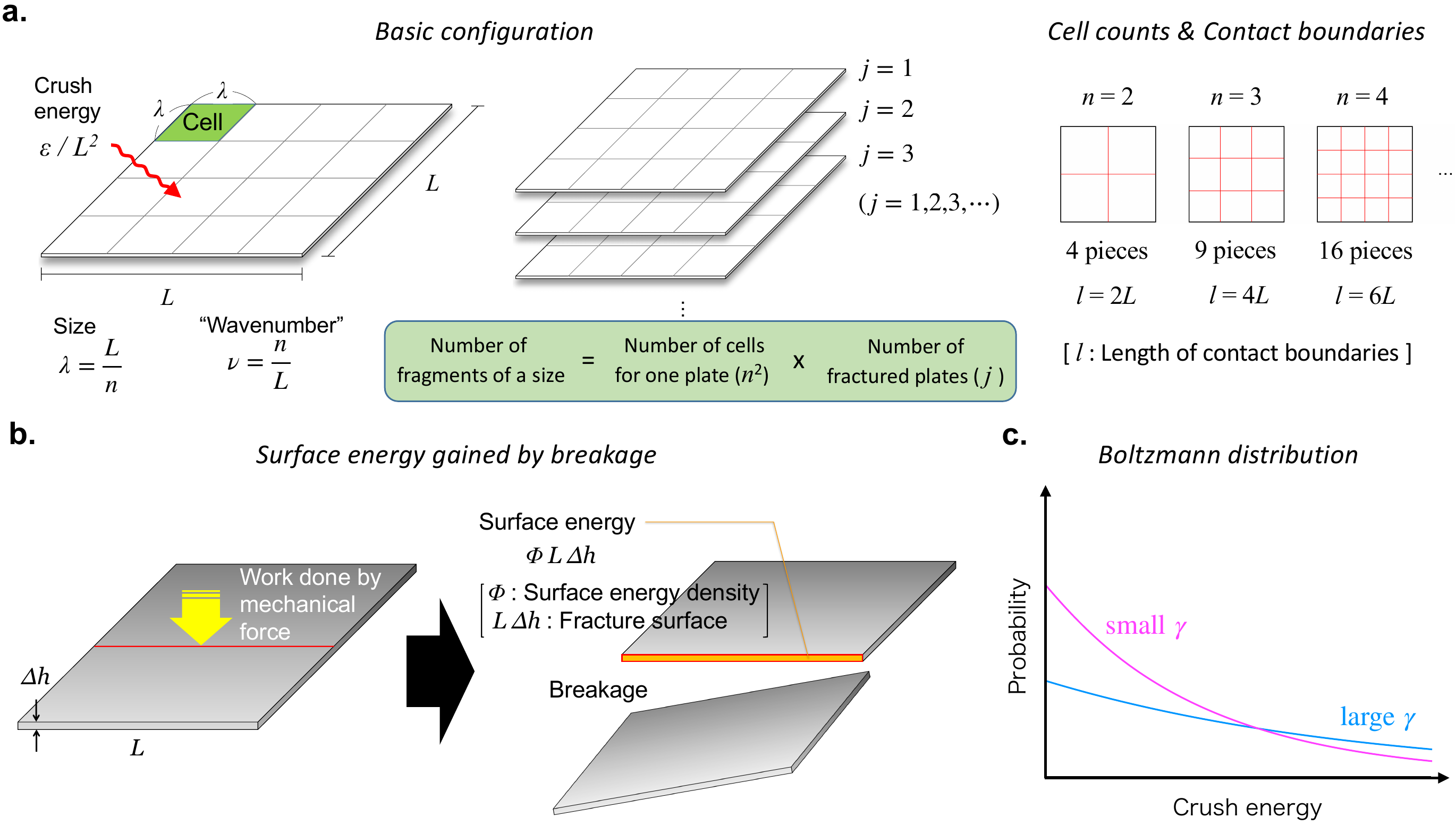}
\caption{\label{merged_fig1}%
 \textbf{Schematic illustration of model configuration.} {\bf a.} Schematic representation of our fracture model.
An idealized plastic plate
with a size of $L \times L$
and a thickness of $\Delta h$
is broken into $n \times n$ equal-sized pieces.
The size of each broken piece is $\lambda=L/n$
(left).
The total length of the contact boundaries
(red lines)
is obviously $l \equiv 2(n-1)L$
(right).
We sum up energy needed to break $j$ plates into
$\lambda$-sized pieces and denote it as $\varepsilon$ (middle).
{\bf b.}
Schematic representation of breakage.
Energy needed to break up the plate is proportional
to the cross-sectional area of the contact boundary
and hence to the length of the fracture.
\hlt{
{\bf c.}
Occurrence probability of crush energy governed by the Boltzmann distribution,
$p(\varepsilon) = e^{-\varepsilon/\gamma}/\gamma$,
which exponentially decreases
for larger crush energy $\varepsilon$.
The e-folding scale is $\gamma$ and hence the probability of larger crush energy becomes larger
as $\gamma$ increases.}%
}
\end{figure*}

The model assumes
that the original plastic piece is a square
plate with a size of $L \times L$ and a uniform thickness of $\Delta h$.
This plate is broken into
$n\times n$ equal-sized cells (Fig.~\ref{merged_fig1}a).
\hlt{In this case, the size of each broken piece is $\lambda=L/n$.}
The number of the fragments of
this particular size is given by $n^2$
multiplied by the number of the original plates (Fig.~\ref{merged_fig1}a).

The fragmentation is assumed to be caused by some mechanical force
and the  required ``crush energy'' is assumed to be proportional to the total length of the fractures.
This assumption may be justified by equating the \hlt{crush} energy to the surface energy of the newly created surfaces (Fig.~\ref{merged_fig1}b) \hlt{, the latter in general equal to the minimal work required in the failure of solid material\citep{Butt2013interfaces, BiswasEA2015}.}
Because the total area of the new surfaces is proportional to the total length of the contact boundaries for a plate (Fig.~\ref{merged_fig1}a, right) and because the surface energy is proportional to the area of the new surfaces\citep{Butt2013interfaces, BiswasEA2015}, the crush energy is proportional to the total length of the fracture.
To produce smaller pieces requires larger energy, which ultimately limits the number of small fragments
as shall be seen below. This is the most significant new element of our model. 

Plastic pieces are fragmented by the action
of waves, winds, or sand under various conditions, for example, on a hard reef or on soft sand,
and therefore energy exerted on the plastic pieces is considered random.
Unless each of these processes is modeled in detail,
however, it is impossible to calculate the probability
distribution of the energy from first principles.
Nor is it possible to infer the probability distribution from field measurements at present.
To make progress, we assume that the Boltzmann distribution,
$p(\varepsilon) \propto e^{-\varepsilon/\gamma}$,
governs the occurrence probability of crush energy
$\varepsilon$
(Fig.~\ref{merged_fig1}c).
This distribution is often assumed
\emph{by default}
when details of the underlying stochastic processes are not known and the distribution is known to be applicable to a wide variety of
situations\citep{Devladar2011,Bouchet2012,VallianatosEA2016,Wu+McFarquhar2018}.
This generality comes from the fact
that it is the distribution that maximizes the entropy under fixed total energy\citep{Kittel1970thermal}; that is, it is the most probable energy distribution under fixed total energy.
According to this probability distribution,
a crush event with a large energy value is less frequent,
consistent with common expectation.
The factor $\gamma$ may be regarded as a representative energy level of
\emph{the environment,}
which represents the aggregate impacts
of the combination of weather conditions (winds, waves, etc.) and the background conditions (hard or soft surfaces, etc.).
We however note that the long-term statistics of the coastal wave height generally shows that a larger wave height occurs with a smaller frequency\citep{Holthuijsen2007, TuomiEA2011, Haselsteiner+Thoben2020}, qualitatively consistent with the Boltzmann distribution.
Nevertheless, our main justification of using this distribution is that our theoretical prediction agrees well with observation as shown below.

The number of the fragments of a particular size is $n^2 j$ for a single crush event
(Fig.~\ref{merged_fig1}a).
Since crush events occur randomly,
the size spectrum is the expected value
of $n^2 j$ as a funcrtion of $\lambda$.
Calculating the expected value
with use of $n=L/\lambda$,
we arrive at the size spectrum
(see ``\nameref{sec:method_theory}" of the Materials and Methods section)
\begin{align}
  \label{eq:size-spectrum-lambda}
  S(\lambda)d\lambda = \frac{A}{\lambda^4}\frac{1}{e^{b/\lambda\gamma}-1}d\lambda,
\end{align}
where $A$ is
an arbitrary positive constant,
\hlt{which is adjusted by the amount of the total mass of plastics to be fractured. Eq.\,\ref{eq:size-spectrum-lambda} is analogous to the Planck's energy spectrum of the black body radiation. See \ref{sec:planck} for the analogy.}

\begin{figure*}[pt]
  \centering\includegraphics[width=1.0\textwidth]{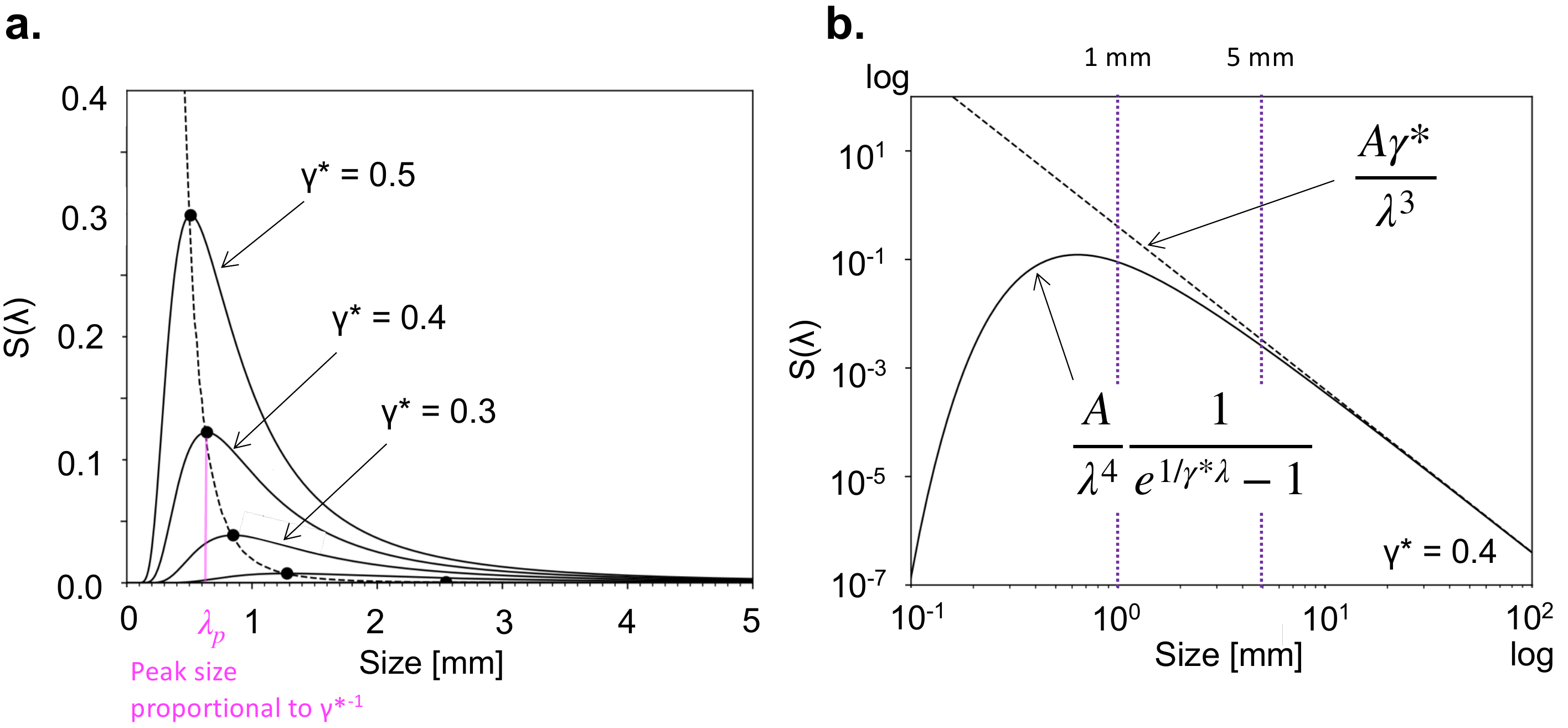}
\caption{\label{fig_theoretical-curve}%
\textbf{Theoretical size spectrum. }
{\bf a.} size distributions expected
from~\eqref{eq:size-spectrum-lambda}
for different values of $\gamma^*$
under a fixed $A$ ($=1.0$).
The dashed curve
connects the peaks of the size distributions.
{\bf b.}
a log-log plot of the same distribution
for $\gamma^*=0.4$ and $A=1.0$ (solid curve).
The dashed line denotes
the power law $A\gamma^*/\lambda^3$,
which the size-distribution
curve asymptotes to for large $\lambda$.
}
\end{figure*}

This size distribution
has a shape skewed to smaller sizes similarly to the observed size distribution of the microplastics,
as shown in
Fig.~\ref{fig_theoretical-curve}.
For a fixed $A$, the size distribution is controlled by $\gamma$ and $b$.
The factor $b$ depends on the plastic material and the size
of the original plate.
The increase of $\gamma$ and the decrease of $b$, however, have the same effect on controlling the size distribution and hence we introduce a new parameter $\gamma^*\equiv \gamma/b$.
This parameter measures
the strength of the mean environmental energy against the strength of the plastic material \hlt{and its inverse provides a characteristic size such that the peak size, $\lambda_p$, is given by $\lambda_p \simeq 0.255{\gamma^*}^{-1}$.}
As this parameter increases, 
thus,
the size at which the maximum of the size distribution occurs
decreases like ${\gamma^*}^{-1}$
and the corresponding maximum value increases like ${\gamma^*}^4$ (Fig.~\ref{fig_theoretical-curve}a).
Furthermore, in the large size limit, i.e., $\lambda\gamma^* \gg 1$,
this size distribution asymptotes to $S(\lambda)d\lambda \sim A{\gamma^*}/\lambda^3 d\lambda$
(Fig.~\ref{fig_theoretical-curve}b),
consistent with the cube power law observed in the mesoplastic range \citep{CozarEA2014plastic}.

The total abundance and mass of the fragments for $\lambda < \Lambda$ ($\leq L$) are obtained, respectively, as
\begin{align}
  N
  &\equiv \int_0^{\Lambda} S(\lambda)d\lambda \approx
   \int_0^\infty S(\lambda)d\lambda
    = \sigma A{\gamma^*}^3, \nonumber\\
  M
  & \equiv
    \int_0^{\Lambda}\!\rho\Delta h \lambda^2 S(\lambda)d\lambda
  = \rho\Delta h A \gamma^* \ln (1 - e^{-1/\gamma^*\!\Lambda})
  \approx \rho\Delta h A \gamma^* \ln (\gamma^* \Lambda),
  \label{eq:integral}
\end{align}
where $\sigma \equiv 2.404$ (\ref{sec:total_abundance}), $\rho$ is the plastic density\hlt{, and $\Lambda$ is the maximum size of plastic fragments}. Both of the above approximations are for $\gamma^*\Lambda \gg 1$,
\hlt{that is, $\Lambda$ being sufficiently larger than the characteristic size $\lambda_p$.}
See \ref{sec:total_abundance} for details.

%
%
\hlt{We fit our model \eqref{eq:size-spectrum-lambda} to an observed size spectrum by adjusting $A$ and $\gamma^*$.
Observed size distributions are presented in various ways: as the number per unit volume of sampled sea water,
as the number per unit surface area of the ocean, as the raw number of collected fragments, etc.
Some studies
normalize the spectrum so that $\int S(\lambda) d\lambda = 1$
\citep{EndersEA2015plastic, KooiEA2019plastics, PoulainEA2018plastics}.
However, note that
the size spectrum \eqref{eq:size-spectrum-lambda} can
take any of these forms by adjusting $A$ and different representations of the same spectra do not affect our analysis because
the differences are
absorbed into $A$.}

We begin by comparing the present theory with the observed size distribution obtained by Isobe et al\cite{IsobeEA2015plastic}.
This size distribution is based on the collection of plastic fragments
sampled around Japan (Fig.~\ref{merged_fig2}a);
this is the largest collection with the highest size resolution to date.
For a precise comparison, we have converted the original size histogram into spectral density.
Details are shown in ``\nameref{sec:method_isobe}" of the Materials and Methods section.

\subsection*{Application to observed data.}
\begin{figure}[pt]
  \centering\includegraphics[width=1.0\textwidth]{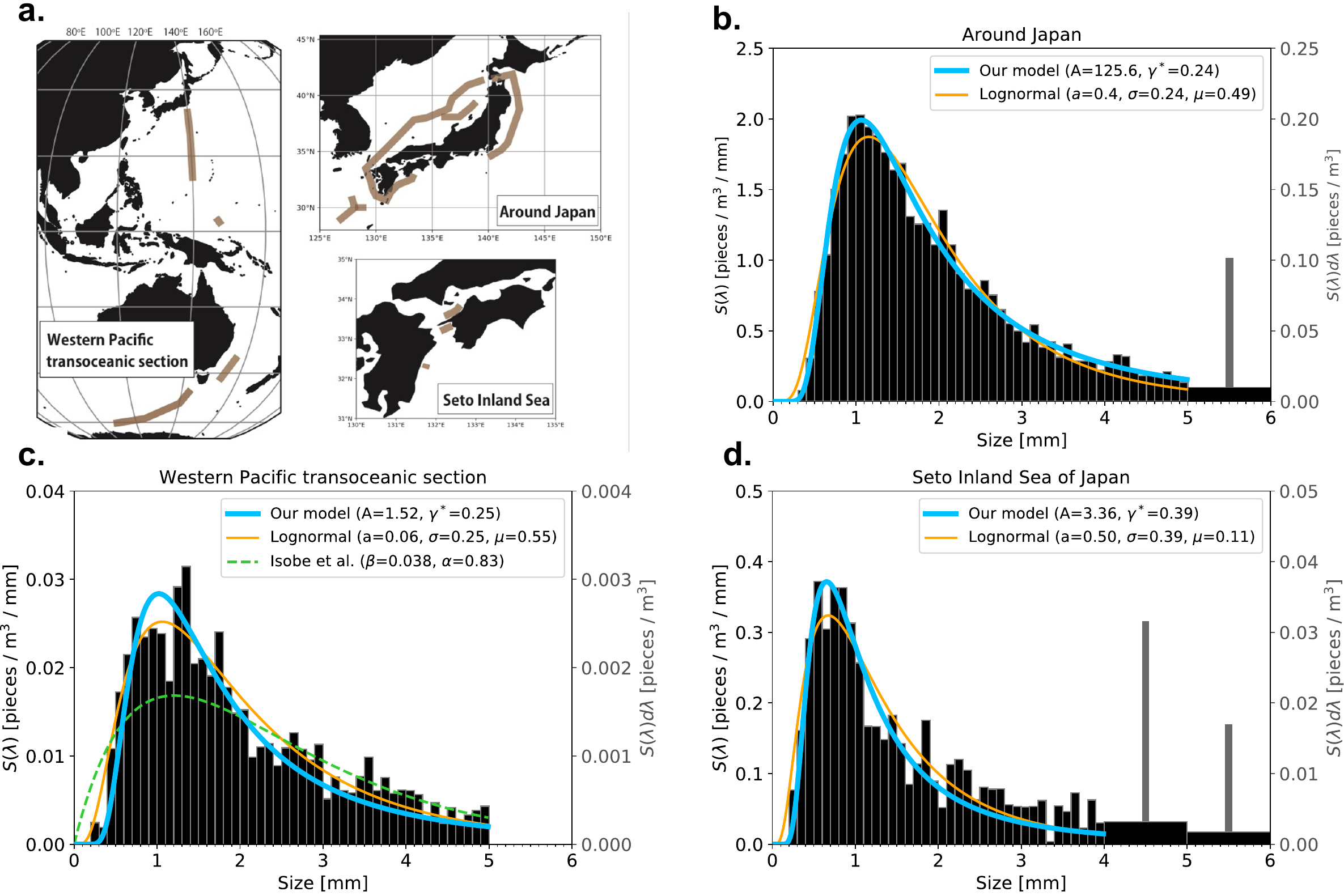}
\caption{\label{merged_fig2}%
\textbf{Comparison of theoretical size spectrum with the observed size distributions of Isobe et al\citep{IsobeEA2014plastic, IsobeEA2015plastic, IsobeEA2019plastic}.}
\textbf{a.} Schematic map of observation stations in Isobe et al\citep{IsobeEA2014plastic, IsobeEA2015plastic, IsobeEA2019plastic}. See their original papers for the detailed locations.
\textbf{b--d.} Theoretical and observed size spectral densities of microplasitcs in the area surrounding Japan,
along a Western-Pacific transoceanic section, and in the Seto Inland Sea.
The black bars display the observed size spectral density (left axis),
and gray bars are the original histogram (right axis) taken from Isobe at al.
In the range $\lambda < 5\,\mathrm{mm}$, the black and gray rectangles
perfectly coincide.
See ``Observed data in Isobe et al'' in the Method section \hlt{for the detailed method of the conversion from the histogram to the spectrum}.
The blue and orange curves denote our model and the lognormal distribution, respectively.
The parameter $A$
is dimensionless ($\times10^{-9}$) in this case.
The green dashed curve in \textbf{c} denotes the empirical curve $\beta\lambda e^{-\alpha\lambda}$ of Isobe et al\cite{IsobeEA2019plastic}. \hlt{The map in {\bf a} is created using Cartopy (version 0.18) (\url{https://scitools.org.uk/cartopy}) embedded in Python and Adobe Illustrator CS6 (version 691) (\url{https://www.adobe.com}).}}
\end{figure}%

Theoretical curve is fitted to the observed size spectrum by adjusting the parameters $A$ and $\gamma^*$
by a least-squares method over
$\lambda < 5\,\mathrm{mm}$. 
The theoretical curve fits the observed size distribution well over the whole microplastic range (Fig.\,\ref{merged_fig2}b)
\hlt{with the relative error, $\Err = 7$\%.
(The relative error is defined by the norm of difference between the theoretical and observed size distributions
divided by the norm of the observed size distribution.)}
For comparison, we also plot a lognormal distribution,
$a\operatorname{LN}(\mu, \sigma^2)$,
\hlt{with an amplification or normalization factor $a$},
where the parameters $a$, $\mu$, and $\sigma$ are determined
by the same least-squares method.
The lognormal distribution is not much different from our model in this size range.

Our model also agrees
well with the observed size distributions in the Western-Pacific transoceanic section \hlt{($\Err=22$\%)} and the Seto Inland Sea \hlt{($\Err=21$\%)} (Figs.\,\ref{merged_fig2}c and~d)
\hlt{with somewhat larger error than around Japan. The observed spectra are not as smooth as that around Japan, suggesting that the samples are not sufficient to give smooth distribution and this may be the reason for the larger error}.
Also, the peak is located at a smaller size in the Seto Inland Sea
than in the other two regions;
this shift is reflected in a larger value of~$\gamma^*$.
The lognormal distribution is somewhat more different
from our model in these two regions
(Figs.\,\ref{merged_fig2}c and~d)
than in the area surrounding Japan (Fig.\,\ref{merged_fig2}b).
The empirical fitting curve used in Isobe et al\cite{IsobeEA2019plastic}
has larger error, despite having the same number of parameters as ours (Fig.~\ref{merged_fig2}c).

\begin{figure*}[pt]
  \hspace*{-0.01\textwidth}%
  \includegraphics[width=1.0\textwidth]{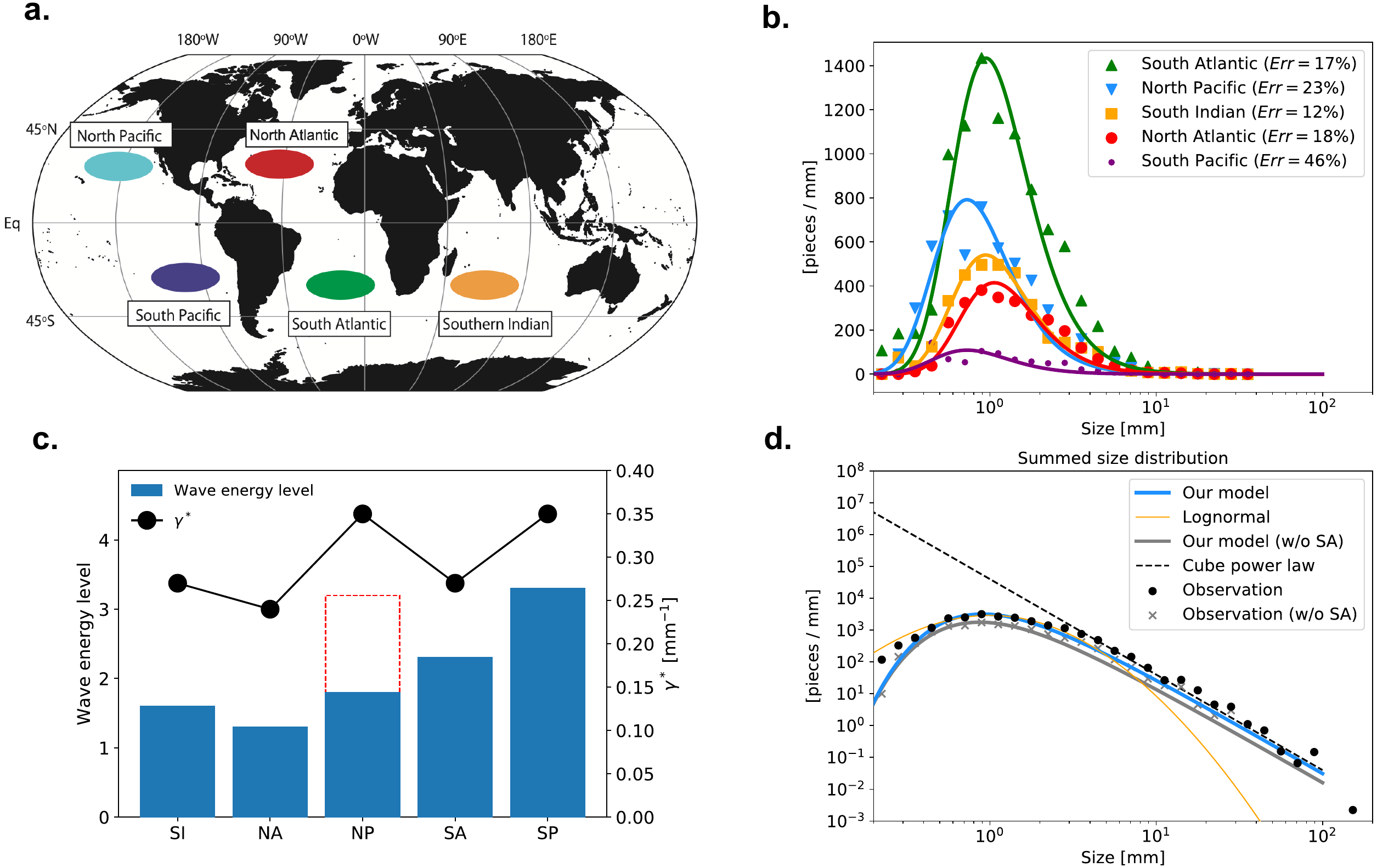}
\caption{\label{merged_fig3}%
\textbf{Application of theoretical model to C{\'o}zar et al\citep{CozarEA2014plastic}'s observational data and comparison between $\gamma^*$ and wave energy.} 
\textbf{a.} Schematic map of observation stations in C{\'{o}}zar et al\cite{CozarEA2014plastic}. See the original paper for the detailed locations.
{\bf b.} Theoretical (curves) and observed (symbols) size spectral density of all samples for each accumulation zone.
The observed data are same as those in their Fig.\,S6, except that the values are expressed as a spectral density (see ``\nameref{sec:method_cozar}" in the Materials and Methods section).
\textbf{c.} Expected wave energy levels \hlt{(no units)} at source regions (rectangles) and $\gamma^*$ (dots) for accumulation zones.
The accumulation zones are denoted by abbreviations
such as SI for South Indian Ocean. The red dashed rectangle shows the wave energy level for the North Pacific Ocean in the case where the contribution from China is removed. See ``\nameref{sec:method_waveenergy}" in the Materials and Methods section for details.
{\bf d.} Sum of the size spectral densities plotted in {\bf b} over the basins for observation (black dots) and our model (blue curve). The orange curve and black dashed line denote lognormal distribution and a cube power law, respectively. The observed data and cube power law line are the same as those in C{\'o}zar et al's Fig.\,S10. The gray dots and curve are the same as the black dots and blue curve, respectively, except that the South Atlantic data is excluded for the sizes less than 30\,mm (the digitizer does not recognize the small spectral values above 30\,mm for the South Atlantic data in C{\'o}zar et al's figure). 
\hlt{The scale of each axis on panels~{\bf b} and~{\bf d} follows that of C{\'o}zar et~al's Figs.\,S6 and~S10, respectively, except that the vertical scale on panel~{\bf b} is shown as spectral density.} \hlt{The map in {\bf a} is created using the same softwares as in Figure\,~\ref{merged_fig2}a.}}
\end{figure*}

We next compare our model with another set of observations
summarized
by C{\'{o}}zar et al\cite{CozarEA2014plastic} (Fig.~\ref{merged_fig3}a)
based on samples collected in
the accumulation zones (``garbage patches'')
over the globe, where microplastics originating from coastal areas tend to converge by ocean currents\cite{LebrentonEA2012plastic,MaximenkoEA2012plastic}.
C\'ozar et al's
method of sampling plastic fragments is similar to Isobe et al's\cite{IsobeEA2014plastic,IsobeEA2015plastic,IsobeEA2019plastic}, but their size distributions include more size bins in the mesoplastic range than Isobe et al's  (See ``\nameref{sec:method_cozar}" of the Materials and Methods section).
Our model generally reproduces well the observed size distributions although
\hlt{it has large relative error in the South Pacific Ocean (Fig.~\ref{merged_fig3}b), where the least number of samples were collected in C{\'o}zar et al.'s study and the spectrum is the least smooth (barely visible in the plot), suggesting that the number of samples is not sufficient}.
\hlt{The unweighted sum of the five size distributions is shown in Figure~\ref{merged_fig3}d
in a log-log form.}
Importantly, the model reproduces the cube power law toward the mesoplastic range (Blue curve in Fig.\,\ref{merged_fig3}d)
as found by C\'ozar et al.
The theoretical curve is not particularly good in the smallest size range,
but this discrepancy reduces
(gray curve and symbols in Fig.\,\ref{merged_fig3}d)
\hlt{when we exclude the South Atlantic Ocean,
which has large difference from the model in the sizes smaller than 0.5\,mm (green curve and symbols in Fig.\,\ref{merged_fig3}b).}
If the plastic pieces sampled in the South Atlantic came from different regions with a wide variety of
$\gamma^*$ values, this discrepancy may be explained  (Figs.\,\ref{fig_reiwa} and~\ref{fig_keio} in \ref{sec:superposition}).
In contrast, the lognormal distribution does not
follow the cube power law
in the large size range (orange curve in Fig.\,\ref{merged_fig3}d)
and this discrepancy exists also in the case without the South Atlantic data (not shown).
\hlt{Despite having one more adjustable parameter, the log-normal distribution fit the data less well than our model.}

The fitting of our model to observations gives a geographical distribution of $\gamma^*$. 
The~$\gamma^*$ value for C{\'o}zar et al's observation is largest in the Pacific Oceans,
smallest in the North Atlantic, and inbetween elsewhere (Fig.~\ref{merged_fig3}c).
Given that the plastics are likely to be fragmented on beaches possibly by ocean waves\citep{Andrady2011plastic},
the value of $\gamma^*$ would represent wave energy on beaches where plastic waste is littered
or comes ashore before fragmented there and washed away into the ocean as micro- or mesoplastics.
A scenario-based numerical experiment (see ``\nameref{sec:method_waveenergy}" of the Materials and Methods section) suggests that the coastal area located along the western boundary of each basin
is likely the major source region
for the plastics in each accumulation zone,
except that a large fraction of the plastics in the South Indian Ocean comes from South East Asia\citep{LebrentonEA2012plastic}.
Compared with a map of categorized wave energy levels along the world coastlines\citep{FairleyEA2020},
the wave energy level in the potential source regions for a basin (Table~\ref{table2}) appears correlated with the~$\gamma^*$ value for the accumulation zone of the basin
in the emission scenario considering impervious area of land (Fig.\,\ref{merged_fig3}c; the results for the other scenario are found in Table~\ref{table2} and \ref{fig_meiji}).
The only exception is the North Pacific:
the North Pacific accumulation zone has a large $\gamma^*$ value in spite of the relatively small wave energy level in the South China Sea,
likely the major source region for the North Pacific.
In fact, a large fraction of the plastic emission from the East China Sea
is likely to come ashore on the southern coast of Japan \citep{LebrentonEA2012plastic},
where the wave energy level is large (Table~\ref{table2}).
If the majority of plastic waste is fragmented there, therefore,
the large $\gamma^*$ value for the North Pacific may be explained (Fig.~\ref{merged_fig3}c).

The interpretation of $\gamma^*$ for Isobe et al's observations is less straightforward
because the source region is less clear. 
The relatively low value of $\gamma^*$ for the area
surrounding Japan (Fig.\ref{merged_fig2}b and \ref{table_ex1})
may be due to the low wave energy in the East China Sea,
which may be the source region of the microplastics\citep{LebrentonEA2012plastic}.
In contrast,
there are clearly multiple source regions
contributing to the Western Pacific transoceanic section
and it is possible that the distribution is a superposition
of distributions with different $\gamma^*$ values (Fig.~\ref{fig_reiwa} in \ref{sec:superposition}).
It is interesting
that the Seto Inland Sea
has a very large $\gamma^*$ (Fig.\ref{merged_fig2}d and \ref{table_ex1}), which may be due to some local conditions
or to the conditions of some remote locations where the microplastics originate.
It is indeed possible that some
microplastics in this region originate from
the Philippine Sea, where wave energy is large \cite{FairleyEA2020},
as previous studies indicate that
some waters from the Philippine Sea are transported into the Seto Inland Sea through the western boundary current\citep{TakeokaEA1993,IsobeEA2010,Nagai+Hibiya2012}.

%
%
\section*{Discussion}

In summary, we derive a theoretical size distribution of micro- and meso-plastics using a statistical mechanical approach.
It assumes that larger ``energy'' is required to break down the original plastic piece into smaller fragments
and that this energy follows the Boltzmann distribution.
This model well reproduces observed size distributions from the micro- to meso-plastics.
In particular, it naturally explains the rapid decrease toward smaller sizes without invoking a removal of smaller fragments.
\hlt{This model is highly idealized,
and extending this model for more realistic fragmentation processes that may be involved
is a future study.}

As pointed out above, C\'ozar et al's South Atlantic size distribution deviates from our theoretical curve
more than the other size distributions in the same dataset.
\ref{sec:superposition} considers theoretical size distributions that would result
if the sample is a mixture of plastic pieces
originating from various source regions with different values of $\gamma^*$,
assuming that each source contributes the same number of plastic fragments
for simplicity.
The resultant superposition deviates from the single-source size distribution
in a similar way that C\'ozar et al's South Atlantic size distribution deviates from
the single-source theoretical distribution
(Fig.~\ref{fig_keio} in \ref{sec:superposition}).
Interestingly,
the value of $1/\gamma^*$ that best fits the mixture
is close to the average of $1/\gamma^*$ values
of the origins in this simple example
(See ``Size distribution" in \ref{sec:superposition}).

Similarly, a mixture
of plastic fragments produced
in various conditions may need
to be considered for interpreting
the observed total plastic mass.
Isobe et al\cite{IsobeEA2019plastic} show that the total mass of the plastics for sizes below $5$\,mm is about $0.3$\,mg\,m$^{-3}$
for their samples in the Western-Pacific transoceanic surveys assuming that $\rho=1000$\,kg m$^{-3}$. To obtain this value from our $M$
formula \eqref{eq:integral} with the optimal values of $A$ and $\gamma^*$ for their dataset and with $\Lambda=5$\,mm, 
we arrive at a value $\Delta h \approx 1$\,mm. 
Obviously this value of $\Delta h$
is too large for the fragmentation model presented in this study because fragments smaller
than $\Delta h$ cannot be produced
by the two-dimensional fragmentation of a
plate with a thickness of $\Delta h$.
This problem may be resolved if
the value of $\Delta h$ obtained like this is in fact an average
of the various thicknesses of the original plastic plates (see ``Total mass" in \ref{sec:superposition}).

\hlt{%
Even ignoring the effect of such a mixture,
our model implicitly
precludes fragments smaller than the lower
limit of the thickness
of original plastic plates,
which is $\sim$100\,$\upmu$m.
In reality, finer microplastics
are observed in the ocean
and on beaches
using different methods from those in Isobe et al.s or C{\'o}zar et al
\citep{EndersEA2015plastic, PoulainEA2018plastics, EoEA2018plastic, Pabortsava+Lampitt2020plastics}.
We hypothesize that these finer fragments are produced by secondary fragmentation of microplastics.
Since the smaller microplastics are closer to three-dimensional shapes than to a plate,
their fracture would be a three-dimensional process.%
}

\hlt{%
The 3-dimensional version of our model (Fig.~\ref{fig_schem_3d-model} in \ref{sec:3d_model}) can be
similarly constructed to the 2-dimensional version (Fig.\,\ref{merged_fig1}a).
The result is such that the model spectrum \eqref{eq:size-spectrum-lambda}
gets one more factor of $\lambda^{-1}$
and $\gamma^* = \gamma/(3L'^3\phi)$, where
$L'$ is the size of the original microplastic cube.
The peak size is at $\lambda_p = 0.201 {\gamma^*}^{-1}$.
See \ref{sec:3d_model} for a complete derivation.
The theoretical curves of the 3-dimensional model are found
to be in good agreement with the observed size spectra of the fine microplastics
with scales of $O$(10\,$\upmu$m) collected near the sea surface\citep{EndersEA2015plastic,PoulainEA2018plastics},
in the upper ocean from the surface to the mesopelagic layer\citep{Pabortsava+Lampitt2020plastics},
and on a beach\citep{EoEA2018plastic} (Fig.~\ref{fig_finemp1} in \ref{sec:3d_model}).
Those theoretical curves are characterized by $\gamma^*$ 
values of 1.68--14.01\,mm$^{-1}$, which are 10--100 times larger than the two-dimensional $\gamma^*$ values of 0.24--0.39\,mm$^{-1}$
for Isobe et al.'s\citep{IsobeEA2014plastic,IsobeEA2015plastic,IsobeEA2019plastic} and C{\'o}zar et al.'s\citep{CozarEA2014plastic} (Figs.~\ref{merged_fig2} and \ref{merged_fig3}).
Note that, however, the energy requirement for the fragmentation is given by $\gamma$.
As an order-of-magnitude calculation, the value of $\gamma = 3L'^3\phi \gamma^*$
from the three-dimensional $\gamma^*$ value
is $O$($10^3$) smaller than the value of $\gamma = 2L^2\Delta h \phi \gamma^*$ from the two-dimensional $\gamma^*$ value under the reasonable assumptions that $L = 100$\,mm and $\Delta h = L' = 1$\,mm and that the surface energy density $\phi$ is the same.
This small $\gamma$ value comes from the smallness of the initial size, $L'$,
which we assume that the three-dimensional fragmentation starts from.
Plastics with smaller initial sizes can be fragmented by smaller energy since the surface energy decreases with the initial size (see Figs.\,\ref{merged_fig1}b in this text and Fig.~\ref{fig_schem_3d-model} in \ref{sec:3d_model}).
}

\hlt{%
Accordingly,
the fate of the marine plastics
may be as follows.
Original plastic pieces deposited
at a beach are fragmented down to millimeter size by
weather phenomena such as waves.
The two-dimensional version of our model intends to predict the spectrum of these fragments.
Those microplastics are further fragmented into finer pieces
by slow weathering, grinding in sand, or other processes.
The three-dimensional version of our model may also explain the resultant size spectrum of the finer fragments.
Both types of microplastics are washed off into the ocean.
Given that a smaller particles are susceptible to the vertical mixing\citep{Kukulka2012plastics,IsobeEA2014plastic,ReisserEA2015plastic,EndersEA2015plastic}, those finer microplastics may be spread over the mixed layer or deeper, while the larger ones may tend to stay at the sea surface. This scenario based on our fracture models has no contradiction to the observed evidence on the spatially varying size distribution of the microplastics so far.
}

\hlt{%
It has been hypothesized that the rapid decrease toward the small size in the observed size distribution at the surface indicates the deep intrusion
of the finer microplastics.
Some numerical simulations incorporating sinking processes
qualitatively support
this scenario\citep{IsobeEA2014plastic,EndersEA2015plastic,IwasakiEA2017plastic},
but such a rapid decrease as observed has not been
\emph{quantitatively} explained by this process yet.
In addition,  
the trouble is that there have not been observations that measure both large ($>300\,\upmu$m) and fine ($<300\,\upmu$m) microplastics
at the same time and there is no data showing continuous spectrum over the entire microplastic range.
There is not much evidence to prove or disprove
this hypothesis.
}

The main purpose of this paper has been to explain the size distribution
of larger microplastics ($>{\sim}300\,\upmu$m) at the surface,
but it has been suggested
that the rapid decrease toward the small size common
to most size distributions may be an artifact attributable to the method of collection or size detection\citep{EndersEA2015plastic,KooiEA2019plastics}.
However, we note that the observed size distributions in Fig.~\ref{merged_fig2}b and d have different peak sizes even though both samples are collected with the neuston nets of the same standards (See ``\nameref{sec:method_isobe}" of Materials and Methods section, and the original articles\citep{IsobeEA2015plastic, IsobeEA2014plastic}).
This result suggests that the peak and the small-size drop are real.

\hlt{In addition, the reduction in the particle count at smallest sizes
would be a physically reasonable feature when available fracture energy is limited.
Indeed, the general grinding needs larger energy to create smaller particles\citep{Berk2018food,Duroudier2016size}, and the size distribution resulting from a progressive grinding process has this feature\citep{BlancEA2020}.}
The power law, which the literature often invokes to explain size distributions,
is usually ascribed to some progressive fragmentation processes.
Such processes may also result in a decrease at small sizes if a limitation of energy is introduced as it is to our model.
When the observed spectrum drops at smaller sizes, the lognormal distribution is also often invoked
and contrasted to the power law in the literature on experiments for plastic fragmentation\cite{TimarEA2010,KishimuraEA2013}.
Quantitatively, the lognormal distribution does not much differ from our model spectrum (Figs.\,\ref{merged_fig2} and~\ref{merged_fig3})
although it does not agree with the power law in the larger-size range (Fig.\,\ref{merged_fig3}d).
Distributions which were compared to the lognormal distribution in the literature may be explained by similar mechanisms to our model.

The present fracture model may open paths toward developing sophisticated numerical simulations for predicting the production and dispersion of the microplastics. There have been numerical simulations to evaluate the spreading microplastics in the ocean\citep{EriksenEA2014plastic, IsobeEA2019plastic,OninkEA2019plastic,SebilleEA2015plastic,WichmannEA2019plastic}.
In such simulations, virtual parcels representing a mass of plastic pieces are released at source regions and advected by ocean currents.
The amount of plastics released has to be either assumed or calibrated so that the resultant mass distribution matches observations.
\hlt{The size distribution, if that information is necessary, is usually assumed in an ad-hoc manner or calibrated against observations.
Given that the microplastics are likely to originate from the weathering of plastic litter on beaches\citep{CorcoranEA2009plastic,Andrady2011plastic,Andrady2017},
their size distribution would depend on weather and wave conditions that would be different for each beach.}
Our size distribution model may be used to estimate
\hlt{the initial size distribution of plastic fragments}
by parameterizing $\gamma^*$ at the beaches as a function of weather conditions such as winds and waves.

%
%
\section*{Materials and methods}

\subsection*{Details of fracture model}\label{sec:method_theory}

\hlt{In the Fracture model section, we only outlined the derivation of our model spectrum and focused on its interpretation; here we show full details of the derivation.
Our model is derived in analogy with
black body radiation; see \ref{sec:planck}
for the analogy.}

In this model, we consider that a square plastic plate with a size of $L \times L$ and a uniform thickness of $\Delta h$ is broken into $n\times n$ equal-sized cells (Fig.\,\ref{merged_fig1}a). The size of each broken piece is given by
$\lambda=L/n$;
we also define an associated
``wavenumber'' $\nu \equiv \lambda^{-1} = n/L$
for convenience.

\hlt{In general, the minimal energy required to fracture
a lump of solid
is given by the surface energy\citep{Butt2013interfaces},
which is proportional to the area of the newly created surface.}
This area in the present model is proportional
to the total length of the contact boundaries (Fig.\,\ref{merged_fig1}a),
which is related to $n$, like $2L(n-1) \approx 2Ln$. With the aid of $\nu=n/L$, the crush energy for a plate can therefore be written as $b\nu$ with a constant $b\equiv 2L^2\Delta h\phi$, where $\phi$ is a uniform surface energy density, and hence the crush energy for $j$ plates
can be written as
\begin{align}
  \label{eq:crushenergy}
  \varepsilon \equiv j b \nu.
\end{align}

We assume that the occurrence probability of crush energy is governed by the Boltzmann distribution characterized by $\gamma$:
\begin{align}
  \label{eq:boltzmann}
  p(\varepsilon) \propto e^{-\varepsilon / \gamma},
\end{align}
which states that a crush event with a large energy value is less frequent, consistent with common expectation.
In statistical mechanics, 
$\gamma$ is temperature in heat bath times the Boltzmann constant (Fig.~\ref{fig_ex1} in \ref{sec:planck}).
This is the analogy to the ``environment'' for microplasitics.

From~\eqref{eq:crushenergy}, the expected value of the crush energy as a function of $\nu$ is
\[
{\langle \varepsilon\rangle}_\nu
=
\sum_{j=0}^\infty jb\nu\, p(jb\nu)
= b\nu {\langle j \rangle}_\nu,
\]
and from~\eqref{eq:boltzmann},
\begin{align}
  \label{eq:bose}
  {\langle j \rangle}_\nu
 = \frac{\sum_{j=0}^\infty j e^{-jb\nu/\gamma}}{\sum_{j=0}^\infty e^{-jb\nu/\gamma}}
  = \frac{1}{e^{b\nu/\gamma}-1},
\end{align}
which is the expected value of the number of fractured plates for each $\nu$.
This formula is called the Bose distribution\citep{Bose1924}.
Since the number of the fragments of a particular size is $n^2 j$ (Fig.\,\ref{merged_fig1}a)
and $n = L\nu$ by definition, the expected number of fragments
is $P(\nu) \propto n^2 \langle j\rangle_\nu \propto \nu^2 \langle j \rangle_\nu$
and therefore,
\begin{align}
  \label{eq:size-spectrum-nu}
  P(\nu) d\nu = A\nu^2\frac{1}{e^{b\nu/\gamma}-1}d\nu,
\end{align}
or converting from $\nu$ to $\lambda$,
we arrive at the size spectrum we already presented in the Fracture model section
\begin{align}
  S(\lambda)d\lambda = \frac{A}{\lambda^4}\frac{1}{e^{b/\lambda\gamma}-1}d\lambda,
  \tag{\ref{eq:size-spectrum-lambda}}
\end{align}
where $A$ is
an arbitrary positive constant.

\subsection*{Observed data in Isobe et al}\label{sec:method_isobe} 
The observed size distributions in Isobe et al are from surveys
around Japan\citep{IsobeEA2015plastic},
in a western Pacific transoceanic section\citep{IsobeEA2019plastic},
and in Seto Inland Sea\citep{IsobeEA2014plastic} (Fig.~\ref{merged_fig2}a).
The first set of samples is from~56 stations around Japan
during the period of July~17 through September~2, 2014 \citep{IsobeEA2015plastic}.
The second is a transoceanic survey at~38 stations across a meridional transect from the Southern Ocean to Japan during 2016\citep{IsobeEA2019plastic};
the stations in Southern Ocean
and the other ones
were occupied
from January~30 to February~4
and
from February~12 to March~2, respectively.
The third dataset is collected at 15 stations in the Seto Inland Sea of Japan
in May--September from 2010 to 2012\citep{IsobeEA2014plastic}.

For all surveys, neuston nets
with mouth, length, and mesh sizes of $0.75\times0.75$\,m$^2$, 3\,m, 0.35\,mm, respectively,
were used to sample small plastic fragments.
The nets were towed near the sea surface around each station for 20\,min
at a constant speed of
2--3\,knots (1--1.5\,m/s).
The numbers of fragments are counted for each size bin with a bin width of 0.1\,mm for microplastics
and 1\,mm for mesoplastics (defined to be $>$5\,mm) between~5 and 10\,mm,
except for the surveys in the Seto Inland Sea,
in which the bin width becomes wider beyond the size of 4\,mm.
The size of a fragment is defined by the longest dimension of its irregular shape.
The concentration of the fragments (pieces per unit volume of sea water)
within each size bin were calculated
by dividing the number of fragments by the water volume measured by the flow meter at each sampling station.
This concentration binned according to size is the observational data used in the present study.
To obtain the numbers,
we have digitized the plots of Isobe et al's
using WebPlotDigitizer version~4.3 (\url{https://automeris.io/WebPlotDigitizer/}).
The original size distributions in the Seto Inland Sea are presented separately for four different areas\citep{IsobeEA2014plastic},
but they are averaged in the present study. 

\begin{figure*}[ht!]
  \centering\includegraphics[width=1.0\textwidth]{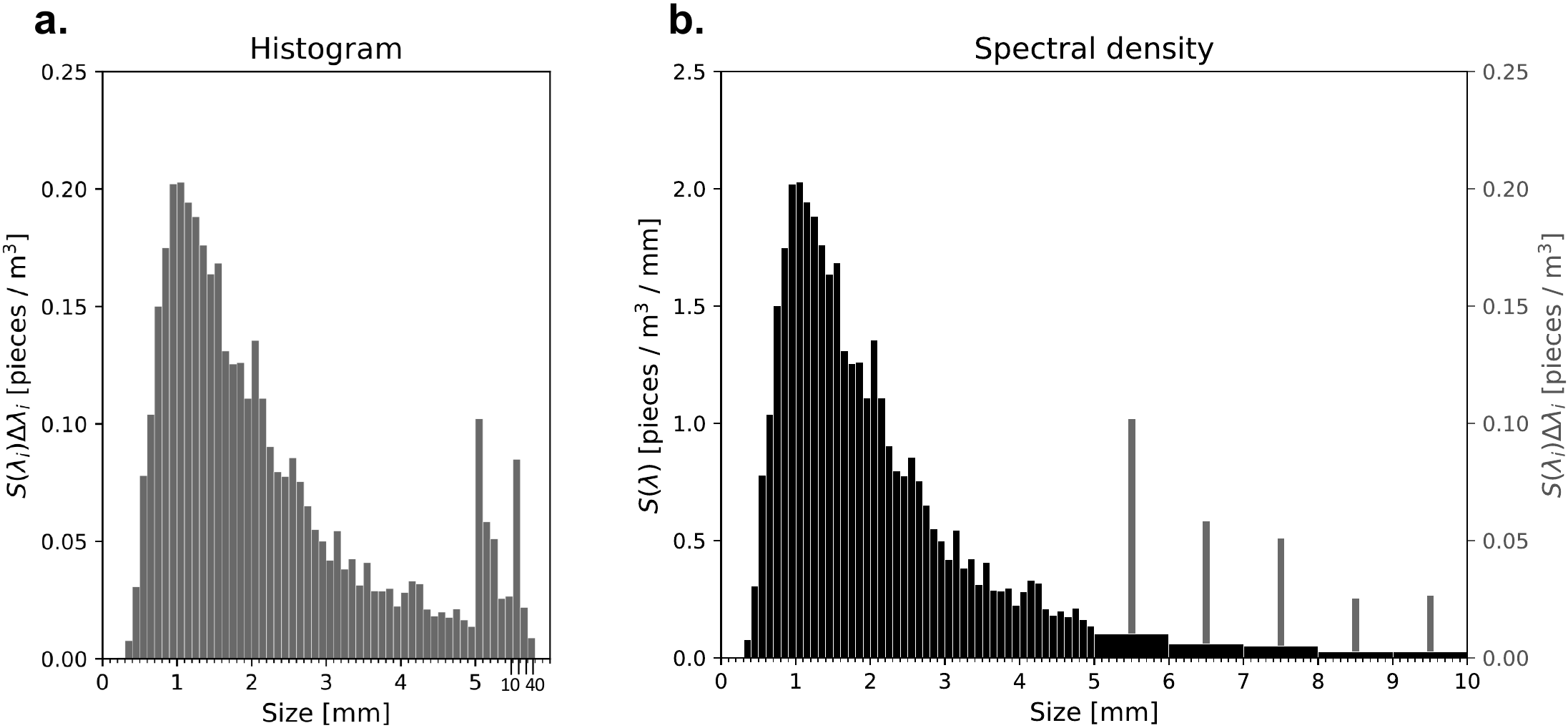}
\caption{\label{fig1}%
\textbf{Size distribution expressed as a histogram ($\bf a$)
and as a spectral density ($\bf b$) from Isobe et al's\cite{IsobeEA2015plastic} observation around Japan.}
The histogram is a replot of Isobe et al's Fig\,2.
We have obtained the data by digitizing the original figure
using WebPlotDigitizer version~4.3 (\url{https://automeris.io/WebPlotDigitizer/}). The size spectral density is indicated by black bars
with its scale shown on the left axis of panel~(b),
and the gray bars on panel~(b)
are the same histogram with its scale on the right axis.
The spectral density is plotted
in such a way that the black bars exactly coincide with the gray ones
for $\lambda < 5\,\mathrm{mm}$.
In panel~(b), sizes larger than 10\,mm are omitted because the spectral values are almost zero there.}
\end{figure*}

Figure~\ref{fig1}a replots one of these
size distributions as an example.
As stated above, the bin width is not uniform:
for this particualr data,
$\Delta\lambda = 0.1\ \mathrm{mm}$
 for $\lambda < 5\ \mathrm{mm}$,
$\Delta\lambda = 1\ \mathrm{mm}$
 for $5\,\mathrm{mm} < \lambda < 10\,\mathrm{mm}$, and
$\Delta\lambda = 10\ \mathrm{mm}$ for $\lambda > 10\ \mathrm{mm}$.
This is the reason that the concentration jumps up
beyond $\lambda > 5\ \mathrm{mm}$.
For comparison with theories,
we introduce a ``size spectral density'' $n_i$
(Fig.\,\ref{fig1}b)
such that
\begin{align}
   \Delta\lambda_i\ n_i = N_i,
\end{align}
where $\Delta\lambda_i$ is the width of the $i$-th bin
and $N_i$ is Isobe et al.'s value for the bin.
By this definition,
each rectangular area of the spectral plot is proportional
to the number of plastic fragments within the bin.
The spectral density is less sensitive to the bin widths,
and in the limit of $\Delta\lambda \to 0$,
it converges to a continuous size spectrum $S(\lambda)$
such that $\int_{\lambda_a}^{\lambda_b}d\lambda\ S(\lambda)$
is the number of fragments between $\lambda_a$ and $\lambda_b$.
All of Isobe et al's size distributions are converted to size spectral densities
in the present study.

\subsection*{Observed data in C{\'o}zar et al}\label{sec:method_cozar}
The plastic samples summarized in C{\'{o}}zar et al\cite{CozarEA2014plastic}
were collected in the accumulation zones\citep{MaximenkoEA2012plastic}
around the world (Fig.~\ref{merged_fig3}a) from December 2010 to July 2011.
Their method of sampling was the same as that of Isobe et al's studies, except for the different mesh size (0.2\,mm at the minimum) and mouth area
($1\times0.5$\,m$^2$) of the neuston net and the different towing durations (10--15 min).
The collected plastic fragments
are classified into bins whose widths increase exponentially
(like $\Delta \lambda_k = c 10^{0.1k}$\,mm, where $k$ is the bin number)
for $0.2\,\mathrm{mm} < \lambda < 100\,\mathrm{mm}$.

C\'{o}zar et al summed the number of plastic pieces over each accumulation zone
without dividing the number by the volume of the sampled sea water
(Fig.\,S6 in their paper).
Their data are digitized and converted to spectral densities in the same way
as described above for Isobe et al's data.
Sizes larger than 40\,mm in the tail of the distribution have had to be omitted
because the numbers are so small there that the data points are too close to
the horizontal axis of the plots for the digitizer to resolve.
This limitation does not significantly influence
the curve fitting because the fitting result is most highly sensitive
to the values of the size spectral density around the peak size.
The present study has also digitized C\'{o}zar et al's Fig.\,S10b
to create a logarithmic plot of the sum of the size distributions
for all the observed accumulation zones (Fig.\,\ref{merged_fig3}d).

\subsection*{Wave energy in source region}\label{sec:method_waveenergy}
We use the Lebreton et al's plastic dispersal simulation result\citep{LebrentonEA2012plastic} to identify
the source regions of the plastic fragments sampled
in the C{\'o}zar et al's accumulation zones.
Lebrenton et al's numerical simulation provides the dispersion of
virtual particles representing a set of plastic fragments originating from land
following the sea-surface currents reproduced
by an oceanic circulation modelling system HYCOM/NCODA\citep{Cummings2005}.
The particles are released at coastal locations on the basis of two scenarios,
the amount of released particles determined as a function of impervious surface area (ISA-based scenario) and of coastal population density (PD-based scenario), respectively.
The former is intended to reflect contributions from major rivers
and the latter from large cities.
Contribution in percentage of each emission region
to the amount of the plastics in the accumulation zones
are summarized in Table~\ref{table2} from Lebreton et al's supplementary data.
For example, the plastic emission from Australia accounts for about 9.5\% of the amount of plastics found in the South Pacific accumulation zone for the ISA scenario.

Next, we estimate wave energy \hlt{level (no units)} at each emission region
from the 6-grade wave energy levels along the global coastlines provided by Fairley et al\citep{FairleyEA2020}.
Each of Lebreton et al's emission regions for each emission scenario
consists of multiple emission points (their supplementary figures~S2--S7).
We select the point with the largest emission within the region,
look up Fairley et al's figure to determine the wave energy level of the point,
and assign the energy level to the region.
If there are multiple points with similarly large energy levels,
we assign the average energy level to the region.
These energy levels are summarized in the ``Energy Level'' column of Table~\ref{table2};
the left and right subcolumns are for the ISA and PD senarios, respectively.
Some of the energy level values are missing in the table because the selected emission points are not covered in
Fairley et al's map.
\begin{table}[tp]
\begin{adjustwidth}{-2.25in}{0in}
  \caption{\label{table2}%
  Expected wave energy levels
  (based on Fairley et al's\citep{FairleyEA2020} map)
  at source regions
  and plastic emissions (percentage) that contribute
  to the five major oceanic accumulation zones
  based on the two terrestrial release scenarios (ISA scenario and PD scenario)
  from Lebreton et al\citep{LebrentonEA2012plastic}.
  The categorization of the source regions is
  from Lebreton et al's supplementary material.
  The values for the ISA and PD scenarios are indicated in the left and right sub-columns of each column, respectively.
  See the Methods section for the details.}
  \includegraphics[width=1.4\textwidth,clip]{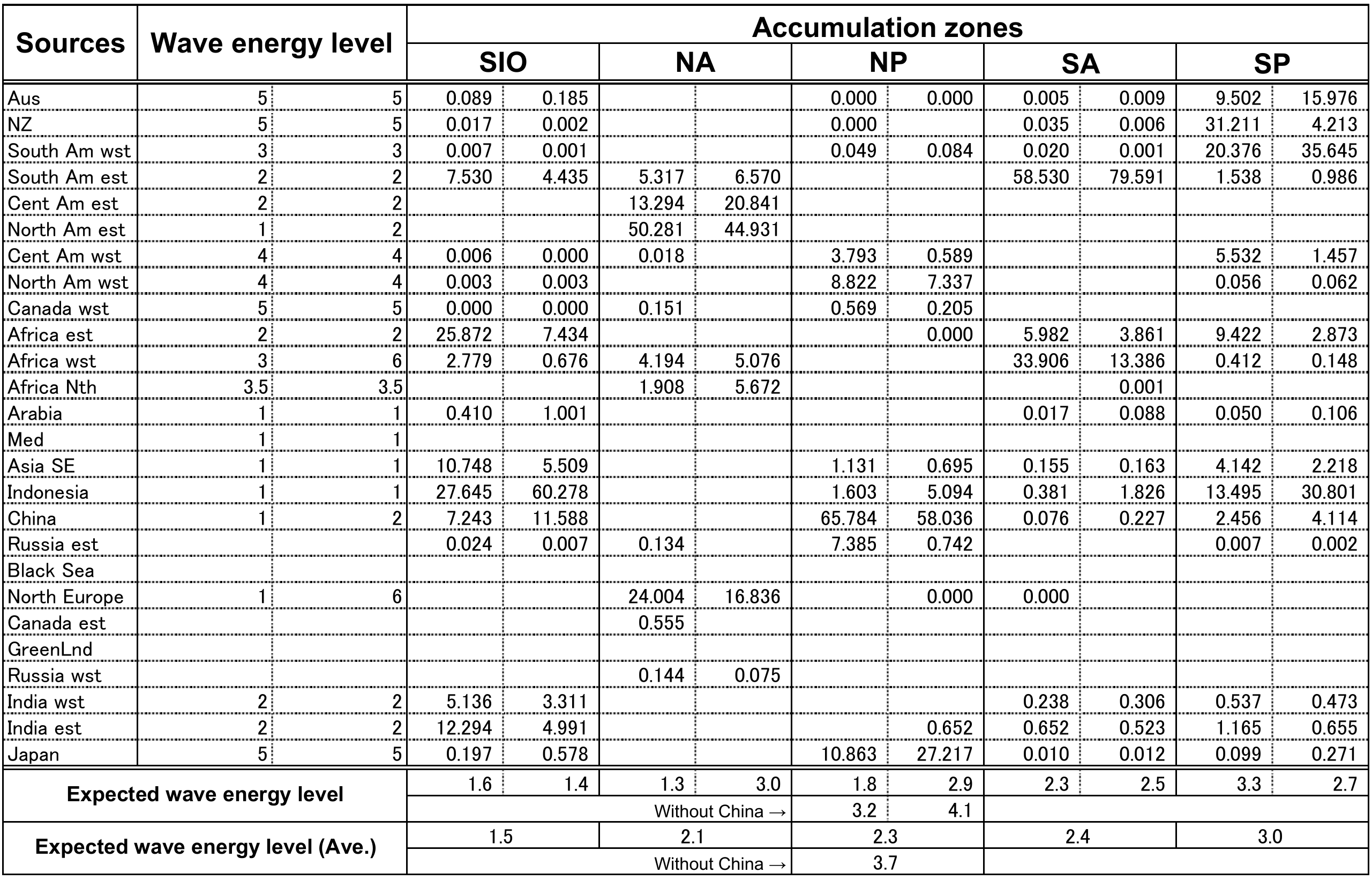}
\end{adjustwidth}
\end{table}
Finally, the expected wave energy level that the plastics found in the accumulation zone experienced for each scenario
is calculated by the contribution-weighted average
$\sum_i E_i f_i/\sum_i f_i$, where $E_i$ and $f_i$ denote the wave energy level and the contribution of the $i$-th source region, respectively.
For the North Pacific,
we calculate the expected wave energy level without contribution from China,
assuming that plastic wastes from China go to the North Pacific accumulation zone via Japan.
The resultant expected wave energy level for each accumulation zone is shown in \ref{fig_meiji}.

\section*{Supporting information}

\paragraph*{S1 Appendix}
\label{S1_Fig}
{\bf Derivation of total abundance.} 

\paragraph*{S2 Appendix}
{\bf Analogy with black body radiation.}

\paragraph*{S3 Appendix}
{\bf Superposition of size distributions.}

\paragraph*{S4 Appendix}
{\bf Three-dimensional model for fine microplastics.}

\paragraph*{S1 Figure}
{\bf Expected wave energy level \hlt{(no units)}
for accumulation zone in Southern Indian Ocean (SIO), North Atlantic (NA), North Pacific (NP), South Atlantic (SA), and South Pacific (SP) (blue bars).}

\paragraph*{S1 Table}
{\bf Optimal $\gamma^*$ for different observation regions.}


\section*{Acknowledgments}
This research was benefited from 
interactions and discussions with Ettore Barbieri, Hiroki Fukagawa, and Yosuke Nakata (alphabetical order).
The observed data were obtained by digitizing the published original figures
with WebPlotDigitizer version~4.3 (\url{https://automeris.io/WebPlotDigitizer/}).

\nolinenumbers

%
%
%
\if0\fi

\bibliography{references}

\include{SupportingInfo}

\end{document}

%% file: SupportingInfo.tex
\appendix
%

\clearpage
\setcounter{page}{1}
\begin{center}
{\bf\large Supporting Information for}\\
{\bf\large ``A model for the size distribution of marine microplastics: a statistical mechanics approach"}\\
{\large Aoki and Furue}
\end{center}
\clearpage

\def\thesection{S\arabic{section} Appendix}
\renewcommand{\theequation}{S\arabic{equation}}
\setcounter{equation}{0}
\renewcommand{\thefigure}{\Alph{figure}}
\renewcommand{\thetable}{\Alph{table}}

\renewcommand{\figurename}{Fig.}
\renewcommand{\tablename}{Table}
\setcounter{figure}{0}
\setcounter{table}{0}
%

\section{Derivation of total abundance\label{sec:total_abundance}}

In this section, we calculate the total number and total mass
of the plastic fragments.
In the main text, the amplitude $A$ is nondimensional
when $S(\lambda)$ is fitted to an observed size spectrum
per unit volume of sea water
or it has the dimension of length cubed
when the observation is a raw size spectrum as
in C\'ozar et al.
Accordingly, the following total number and mass
are regarded as per unit volume of sea water or raw
depending on which type of size spectrum $S(\lambda)$ denotes.

A transformation of variables
$\nu'=\nu/{\gamma^*}$ in~\eqref{eq:size-spectrum-nu} leads to
\begin{align}
  \label{eq:size-spectrum-nudash}
  S(\nu)d\nu = 
  A{\gamma^*}^3\nu'^2\frac{1}{e^{\nu'}-1}d\nu'.
\end{align}
The total number of plastic fragments
over $0 < \lambda < \Lambda$ ($\leq L$)
can then be written as
\begin{equation}
   N  \equiv \int_{1/\Lambda}^\infty S(\nu)d\nu
     =  \int_{1/\gamma^*\!\Lambda}^\infty \frac{A{\gamma^*}^3\nu'^2}{e^{\nu'}-1}d\nu'
\end{equation}
and with
this familiar formula
$
  (e^{\nu'}-1)^{-1} = \sum_{j=1}^{\infty} e^{-j\nu'},
$
\begin{subequations}
\begin{align}
 N &= 
  A\gamma^{*3} \sum_{j=1}^{\infty} 
  \int_{1/\gamma^*\!\Lambda}^\infty \nu'^2 e^{-j\nu'} d\nu' \nonumber\\
  & = A\gamma^{*3}\left[
  2\operatorname{Li}_3(e^{-1/\gamma^*\!\Lambda})
  +2\left(\frac{1}{\gamma^*\!\Lambda}\right)
  \operatorname{Li}_2(e^{-1/\gamma^*\!\Lambda})
  -\left(\frac{1}{\gamma^*\!\Lambda}\right)^2
  \ln(1-e^{-1/\gamma^*\!\Lambda})
  \right],
  \label{eq:N-integral-exact}
\end{align}
where
\[
  \operatorname{Li}_s(z) \equiv \sum_{j=1}^{\infty}\frac{z^j}{j^s}.
\]
When $\Lambda\gg\gamma^{*-1}$,
\begin{equation}
N \approx  A\gamma^{*3} 2\operatorname{Li}_3(1)
  = \sigma A\gamma^{*3},
  \label{eq:N-integral-approx}
\end{equation}
where $\sigma \equiv 2.404$,
because $\operatorname{Li}_3(1) = \sum_{j=1}^{\infty}j^{-3} \simeq 1.202$
(known as Ap\'{e}ry's constant;
see \url{https://oeis.org/A002117}).
\end{subequations}
This approximation is equivalent to $ \int_{1/L}^\infty S(\lambda)d\lambda
\approx
\int_0^\infty S(\lambda)d\lambda$.
This approximation is a natural one
when $\Lambda\sim L$
because
$1/\gamma^*\!\Lambda \sim 1 / \gamma^* L
= 2L\Delta h\phi/\gamma$
and this factor is therefore the ratio of the surface energy
$L \Delta h \phi$ to the mean environmental energy $\gamma$.
We naturally assume that $L \Delta h \phi \ll \gamma$ because otherwise not many small fragments would be generated.

Similarly, the total mass of plastics is
\begin{equation}
  M \equiv \int_0^{\Lambda}\rho\lambda^2\Delta h\,S(\lambda)d\lambda,
\end{equation}
where $\rho$ is the mass density of the plastic material and $\Delta h$ is the thickness
of the original plate.
After similar transformations as above,
\begin{subequations}
\begin{align}
  M = A\gamma^*\rho \Delta h
  \sum_{j=1}^{\infty}\frac{e^{-j/\gamma^*\!\Lambda}}{j}
  &=-A\gamma^*\rho \Delta h\ln(1-e^{-1/\gamma^*\!\Lambda})
    \label{eq:M-integral-exact} \\
  &\approx A \gamma^*\rho \Delta h \ln(\gamma^*\!\Lambda),
    \label{eq:M-integral-approx}
\end{align}
\end{subequations}
using the same approximation, $\gamma^*\!\Lambda \gg 1$, as for $N$.
Unlike $N$,
$M$ depends on $\Lambda$ even when $\gamma^*\!\Lambda \gg 1$
because
the contribution of larger plastic pieces
is significant to $M$ whereas it is negligible
to $N$.
\clearpage

\section{\hlt{Analogy with black body radiation}}\label{sec:planck}

\hlt{Our size spectrum is derived
in analogy with black body radiation.
Here we outline the derivation of the wavenumber
spectrum of black body radiation and
discuss the analogy.
Since the derivations of the Boltzmann distribution and Planck's spectrum below are standard,
we do not cite references there.
See the literature on statistical mechanics for details (e.g., Kittel and Kuroemer 1980\citep{Kittel1970thermal}).}

\paragraph{\hlt{Boltzmann distribution.}}
\hlt{%
Consider a large isolated system (heat bath)
and a small subsystem,
and classify the states which the subsystem
can take by their energy value $E$\@.
Assume that the subsystem takes
a state with $E$ at a probability
of $p(E)$.
Then we calculate the probability distribution that maximizes
the entropy of the subsystem,
\begin{align}
    S = -k\int_0^\infty [p(E)\ln p(E)]\,\Omega(E)\,dE,
    \label{eq:definition-of-S-continuous}
\end{align}
where $\Omega(E)$,
known as the ``density of states,''
is the number
of states with energy value $E$
in the subsystem,
under the constraints
that $\int_0^\infty p(E)\,dE = 1$
and that the expected value of
energy
\begin{equation}
\langle E\rangle =
\int_0^\infty E\,p(E)\,\Omega(E)\,dE
\label{eq:expected-energy-value}
\end{equation}
is given.
The solution is
\begin{align}
    p(E) = \frac{e^{-\beta E}}{Z},
   \quad\quad \text{where}\quad
    Z\equiv\int_0^\infty e^{-\beta E}\,\Omega(E)\,dE.
    \label{eq:boltzmann-continuous}
\end{align}
The variable $Z$ as a function 
of $\beta$ is called the ``partition function''.%
}

\hlt{%
This probability distribution
is known as the Boltzmann distribution.
The variable $\beta$,
which enters the solution because
of the energy constraint,
is related to the temperature
of the system through the thermodynamic relation
\[
\frac{1}{T}
= \frac{\partial S}{\partial \langle E\rangle} = \beta k.
\]
For the second equality,
we have used
\eqref{eq:definition-of-S-continuous}--\eqref{eq:boltzmann-continuous}
to calculate the derivative.
We replace $\beta$ with $k T$ in what follows.%
}

\paragraph{\hlt{Black body radiation.}}

\hlt{Consider a mass of material, a ``black body'',
which is in thermal equilibrium and assume that there is a vacuum
cavity within it.
Photons are emitted from the black body into the vacuum cavity and absorbed by the opposite wall.
The energy of the photons obeys the energy probability distribution of the black body,
which is assumed to be the Boltzmann distribution.
Planck further assumed that the the energy of a photon with frequency $\omega$
can take only a value which is an integral multiple of the unit $\hslash\omega$, where $\hslash$ is a universal constant.
The energy of the photons in the cavity, accordingly, takes the form of
\begin{align}
    \label{eq:energy_photon}
    \varepsilon(j,\omega) = j \hslash\omega.
\end{align}
Since~\eqref{eq:energy_photon} obeys
the Boltzmann distribution,
we can convert the probability 
of photon energy as a function 
of $j$ with $\omega$ regarded as a parameter
using~\eqref{eq:energy_photon}
and~\eqref{eq:boltzmann-continuous}.
The result is
$
p(j;\omega) = e^{-jh\omega/kT}/\sum_j e^{-jh\omega/kT}.
$
In this case, the expected energy can be calculated as
$\langle E \rangle = \sum_j j\hslash\omega\,p(j;\omega)$
and hence
\begin{align}
    \langle E \rangle (\omega) = \frac{\hslash\omega}{e^{\hslash\omega/kT}-1}
  \label{eq:E-omega}
\end{align}
for each frequency.
Also, this energy divided by
$\hslash\omega$ provides the expected number of photons
\begin{align}
    \label{eq:bose_planck}
    \frac{1}{e^{\hslash\omega/kT}-1}.
\end{align}
This formula is known as the Bose distribution.}

\hlt{
Finally, we convert
the energy
distribution
\eqref{eq:E-omega}
to the frequency spectrum of energy
considering the dimensionality of
the space.
The frequency interval
$(\omega, \omega + d\omega)$
includes $\omega^2/\pi^2 c^3$ modes of the wave per unit
volume of the three-dimensional cavity.
Since the energy spectrum is 
the product between the number of modes
included in the interval
$(\omega, \omega + d\omega)$
and the expected value of the energy at $\omega$,
which is given by~\eqref{eq:E-omega},
the result is
Planck's energy spectrum
\begin{align}
  \label{eq:planck}
  \frac{\hslash}{\pi^2c^3}\frac{\omega^3}{e^{\hslash\omega/kT}-1}.
\end{align}%
}

\begin{figure*}
  \centering\includegraphics[width=0.5\textwidth]{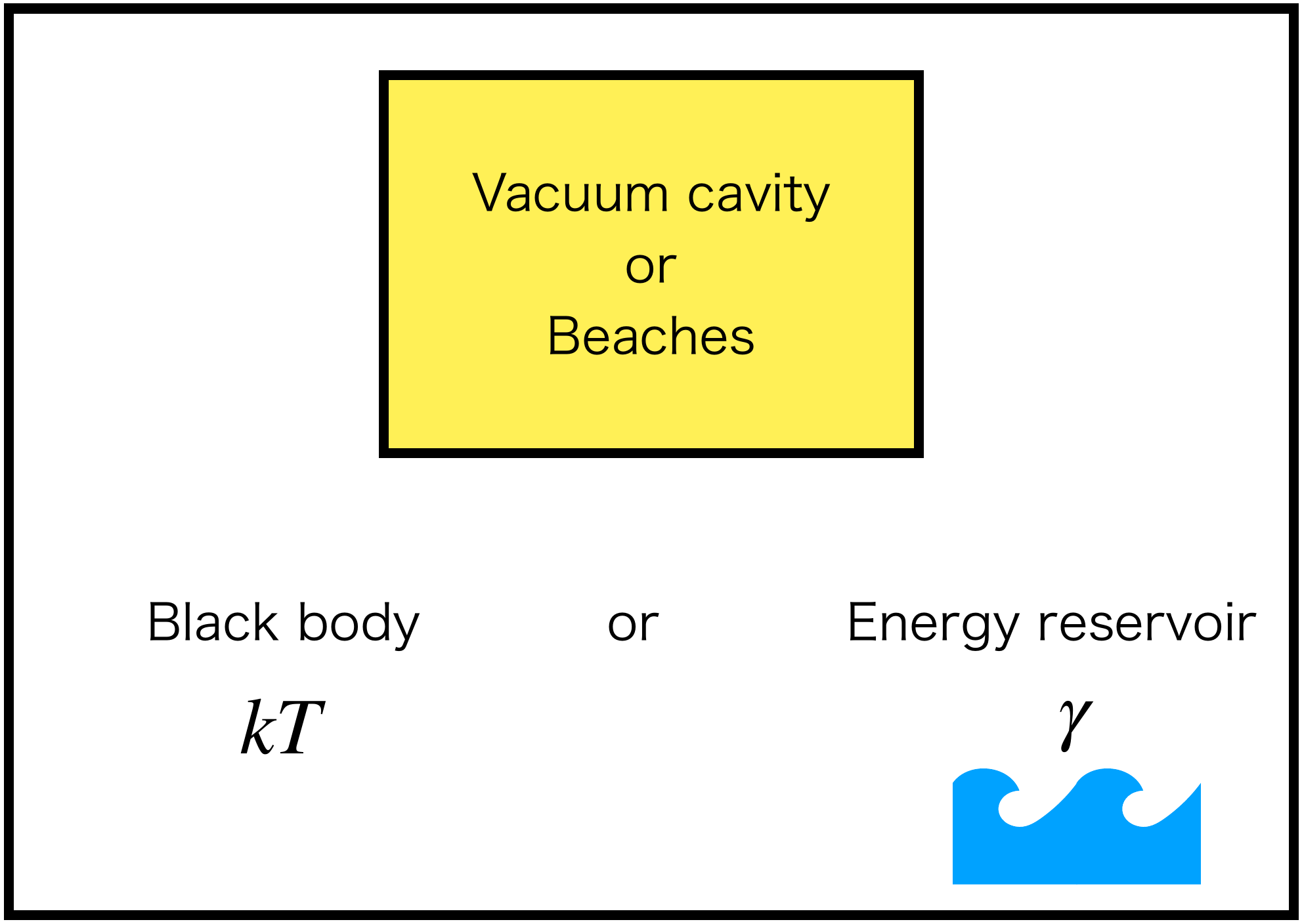}
\caption{\label{fig_ex1}%
\hlt{Analogy between black-body radiation and microplastics.
The yellow area represents the vacuum cavity where electromagnetic radiation occurs
or the beaches where microplastics are produced.
The white area represents
the blackbody characterized by $kT$,
or the energy reservoir (winds, waves, and so on)
characterized by $\gamma$.}
}
\end{figure*}

\paragraph{\hlt{Analogy.}}
\hlt{%
By analogy,
the photons correspond to
the fragmented plastic litters on beaches,
and the black body to the environment (the weather and wave conditions)
(Fig.\,\ref{fig_ex1}).
The mean energy of the black body, $kT$,
corresponds to the mean crush energy, $\gamma$, of the environment.
Further, the crush energy is expressed as $\varepsilon = jb\nu$ (Eq.~\ref{eq:crushenergy}) on
the basis of the necessary surface energy.
This expression corresponds to~\eqref{eq:energy_photon} in Planck's theory.
Because of this analogy,
the Boltzmann distribution (Eq.~\ref{eq:boltzmann} with Eq.~\ref{eq:crushenergy})
and the Bose distribution (Eq.~\ref{eq:bose}) in our model
are formally the same as
those of Planck's theory.}

\hlt{%
Our size spectrum (Eqs.\,\ref{eq:size-spectrum-lambda} and~\ref{eq:size-spectrum-nu}) is analogous to Planck's \eqref{eq:planck}. 
As compared to~\eqref{eq:size-spectrum-nu},
Planck's formula can be obtained formally
if we take $A \to \hbar/\pi^2c^3$ and $\gamma\to kT$, $\nu \to \omega$ after multiplying the right-hand side of~\eqref{eq:size-spectrum-nu}
by $\nu$.
The last multiplication is merely due to the dimensionality of the space:
the numerator of our formula becomes also proportional to $\nu^3$
if the fragmentation of plastics is three-dimensional (see below).
The corresponding wavelength (size) spectrum is obtained from the relation $\lambda = c/2\pi\omega$
for Planck's and $\lambda = 1/\nu$ for our spectrum.%
}

\hlt{%
The transition from~\eqref{eq:E-omega}
to~\eqref{eq:planck} does not have
a close analogy with the plastic model.
As we have seen, Planck's energy distribution is essentially
\begin{align*}
  (\text{number of modes $\propto \omega^2$})
    \times
    (\text{energy of a photon $\hbar\omega$})
\times
   (\text{Bose distribution})
\end{align*}
On the other hand, our plastic ``wavenumber'' spectrum is essentially
\begin{align*}
  (\text{number of fragments $\propto \nu^2$})
\times
  (\text{Bose distribution}),
\end{align*}
where the Bose distribution describes the expected number of original plastic pieces (plates) which are fragmented.
If the original plastic piece is a cube and its fragmentation is three-dimensional,
the number of fragments will be proportional to $\nu^3$ and the functional form of 
the plastic ``wavenumber'' distribution with respect to $\nu$ will be exactly the same as Planck's (See \ref{sec:3d_model}).%
}

\clearpage
\section{Superposition of size distributions}\label{sec:superposition}
\paragraph{Size distribution.}
We explore how the size distribution is modified
if multiple source regions with different parameters contribute.
Here we assume that each source region contributes
the same number of plastic fragments ($N$),
which gives $A$ in~\eqref{eq:size-spectrum-lambda}
as a function of $\gamma^*$:
\[
A = \frac{N}{\sigma\gamma^{*3}}
  = \frac{N}{2.404 \gamma^{*3}}
\]
according to~\eqref{eq:integral}.
In this case, the size spectrum~\eqref{eq:size-spectrum-lambda}
can be written as
\[
S^*(\lambda; a, b^*)
\equiv a \frac{b^{*3}}{\lambda^4}\frac{1}{e^{b^*/\lambda}-1},
\]
where $a$ is a nondimensional constant and $b^* \equiv b/\gamma = 1/\gamma^*$.

\begin{figure*}
  \centering\includegraphics[width=0.8\textwidth]{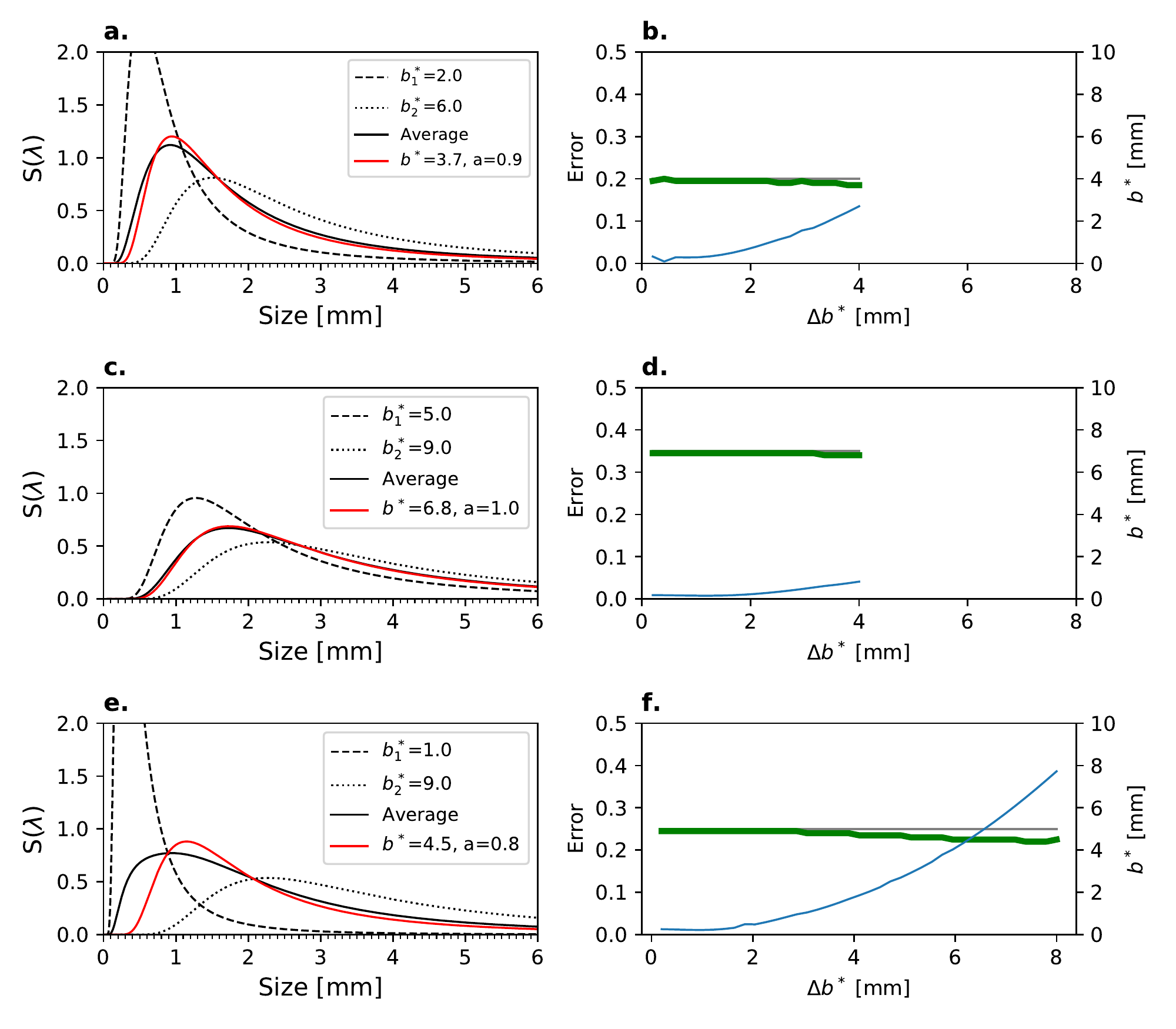}
\caption{\label{fig_reiwa}%
Superposition of size distributions
with different values of~$b^*$ ($=1/\gamma^*$)
ranging from $b^*_c - \Delta b^*/2$
to $b^*_c + \Delta b^*/2$
for
\textbf{(a)}
$b^*_c = 4$\,mm and $\Delta b^* = 4$\,mm,
\textbf{(c)}
$b^*_c = 7$\,mm and $\Delta b^* = 4$\,mm, and
\textbf{(e)}
$b^*_c = 5$\,mm and $\Delta b^* = 8$\,mm.
The dashed and dotted curves on the left panels are
$S(\lambda; 1, b^*_1)$ and $S(\lambda; 1, b^*_2)$, respectively,
where $b^*_{1,2} \equiv b^*_c \pm \Delta b^*/2$.
The solid black curve is $S(\lambda; 1, b^*)$
averaged from $b_1^*$ to $b_2^*$.
The red curve indicates the best-fit size spectral density $S(\lambda; a_{\text{opt}}, b^*_{\text{opt}})$
to the average.
The right panels (\textbf{b,d,f})
show the fitting error (cyan)
and the optimal $b^*$ of the best-fit curve (green)
as a function of $\Delta b^*$,
where $b^*_c$ and the maximum value of $\Delta b^*$ are the same as in the corresponding left panel.
The error is defined as the ratio of the norm of the difference between the average and best-fit curves to the norm of the average.
The thin horizontal gray line denotes $b^*_c$.
}
\end{figure*}

We next calculate the average of
$S^*(\lambda; 1, b^*)$
from $b^*_c - \Delta b^*/2$ to $b^*_c + \Delta b^*/2$.
The order of magnitude of $b^*$
is known
because the value of $\lambda$ that gives the peak of the size spectrum
is $O(b^*)$  (it can be shown that it is approximately $0.255 b^*$ from Eq.~\ref{eq:size-spectrum-lambda} and this $\lambda$ value is constrained by observations).
The average is calculated numerically changing
$b^*$ at an interval of 0.1 mm.
For an illustration,
and we look at three cases with
$(b^*_c, \Delta b^*)
= (4\,\mathrm{mm}, 4\,\mathrm{mm}),
(7\,\mathrm{mm},4\,\mathrm{mm}),
(5\,\mathrm{mm},8\,\mathrm{mm})$,
and plot the results in  Figs.\,\ref{fig_reiwa}a, \ref{fig_reiwa}c, and \ref{fig_reiwa}e, respectively.
The solid black curve plots the averaged $S^*$;
the dashed and dotted curves plot $S^*(\lambda; 1, b^*_1)$ and $S^*(\lambda; 1, b^*_2)$, where
$b^*_1 \equiv b^*_c - \Delta b^*/2$ and
$b^*_2 \equiv b^*_c + \Delta b^*/2$.
%
%
\begin{figure*}
  \centering\includegraphics[width=0.6\textwidth,clip]{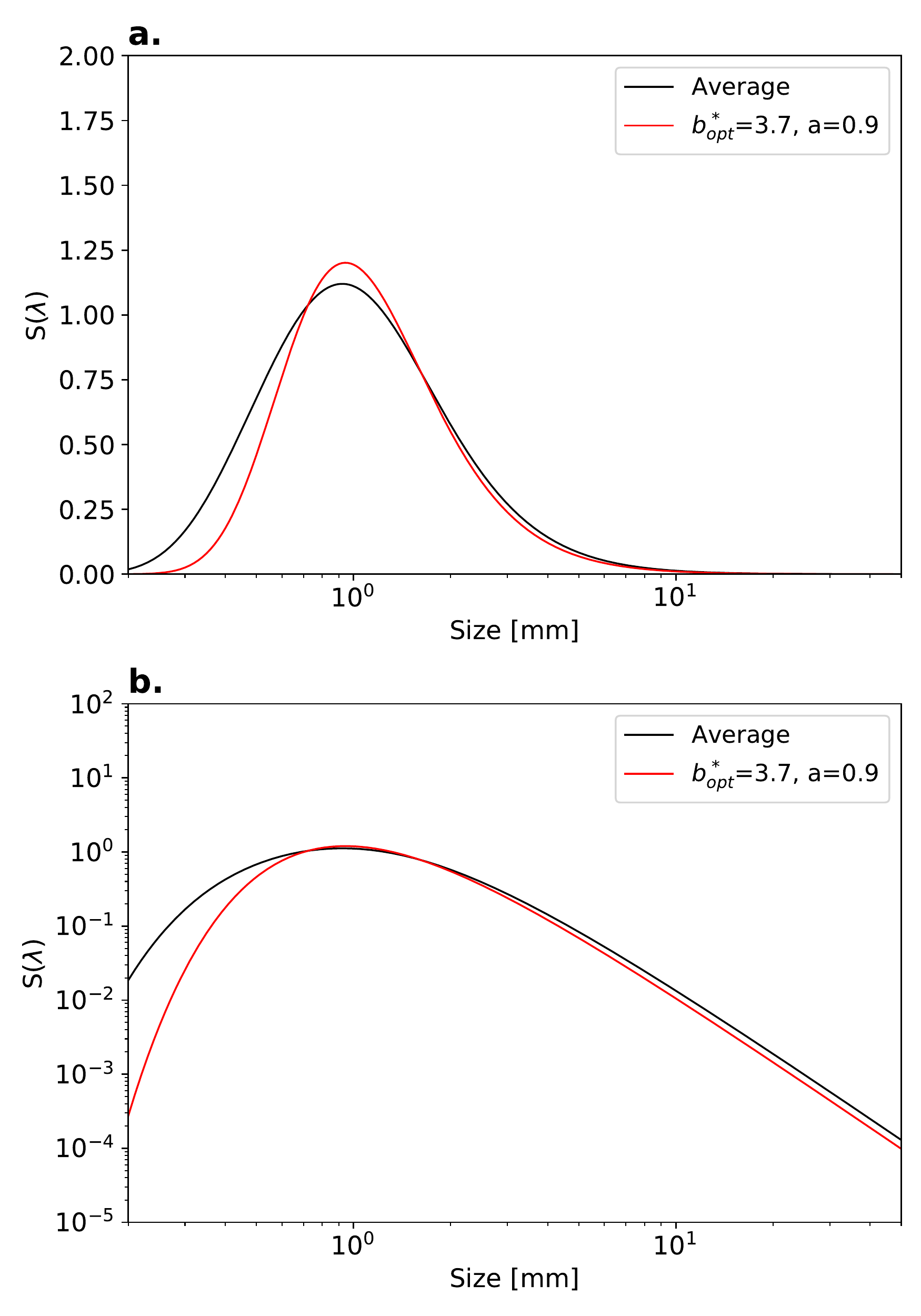}
\caption{\label{fig_keio}%
Size spectral density averaged over $b^*$ (black) and the best-fit curve (red) with optimal
values for $b^*$ and $a$.
Both curves are the same
as the black and red curves of Fig.\,\ref{fig_reiwa}a except the horizontal axis
(\textbf{a}) and both axes (\textbf{b}) are logarithmic in this figure.}
\end{figure*}

We then fit $S^*(\lambda; a, b^*)$ to the average profile
by adjusting $a$ and $b^*$, which is the red curve.
This is to simulate the fitting of our theoretical curve to
an observation which may be a mixture of plastic pieces from different origins.
Compared to the ``pure'' profile (red curve),
the peak of the average profile (black curve)
shifts leftward, the peak value is lower, 
and the values are larger in the smallest size range.

The right panels of Fig.\,\ref{fig_reiwa} plot
the error
(cyan curve)
of the fitting of the pure profile
to the average as a function of $\Delta b^*$
with the same $b^*_c$ value as in the respective left panel,
which corresponds to the maximum value of $\Delta b^*$
of the right panel.
As expected, the fitting error grows with $\Delta b^*$.
The green curve plots the optimal $b^*$ as a function
of $\Delta b^*$.  The optimal $b^*$ changes little
and stays close to $b^*_c$ (thin gray line),
indicating that the value of $b^*$ ($=1/\gamma^*$ by definition)
obtained by fitting observations is close to its average value.

Fig.\,\ref{fig_keio} plots the average
and optimal profiles from Fig.\,\ref{fig_reiwa}a
but with the horizontal axis logarithmic (panel~a)
and with both axes logarithmic (panel~b).
The difference between the two curves 
is qualitatively similar to the difference between the observed
and the best-fit theoretical curves
for C\'ozar et al's South Atlantic
data in Fig.~\ref{merged_fig3}d.

\paragraph{Total mass.}
Here we explore the impacts of superposition
on the total mass.
Suppose that the observed size distribution
is a superposition of different distributions
with different values of $A$, $\gamma$,
$\phi$, $L$, and $\Delta h$ (See Fig.~\ref{merged_fig1}).
We denote those parameter values
for each distribution by $A_k$, $\gamma_k$, etc.~for
$k = 1,\ldots,K$.
Assume that the shape of the superposition
is similar to a ``pure'' distribution
as in Figs.\,\ref{fig_reiwa}a and~\ref{fig_reiwa}c.

By fitting our model spectrum to the observed,
we obtain optimal values for~$b^*$ and~$A$\@.
Because the size distribution is similar to the
corresponding ``pure'' distribution,
\eqref{eq:M-integral-exact}
or 
\eqref{eq:M-integral-approx}
should give an accurate total mass.
In the main text, we used this approach
to infer the value of $\Delta h$ so that the calculated total mass agrees with the observed.
This approach can be formulated by,
if we use the approximate form~\eqref{eq:M-integral-approx} for simplicity,
\[
\frac{\rho\Delta h A}{b^*}
   \ln(\Lambda/b^*)
=
\sum_{k = 1}^K
\frac{\rho_k A_k}{c_k}
   \ln\frac{\Lambda}{c_k \Delta h_k},
\]
where $c_k \equiv b^*_k / \Delta h_k = 2L_k^2\phi_k/\gamma_k$.
The values of $A$ and $b^*$ on the left-hand side are
those obtained by fitting the observed distribution
and $\Delta h$ on the left-hand side is the inferred value.
Therefore, the inferred $\Delta h$ is an ``average''
of $\Delta h_k$'s
in the sense that
\[
\Delta h =
\frac{b^*}{\rho A \ln(b^*/\Lambda)}
\sum_{k = 1}^K
\frac{\rho_k A_k}{c_k}
   \ln\frac{c_k \Delta h_k}{\Lambda}.
\]

Obviously, the result depends on the parameters $A_k$, $\gamma_k$, etc.
If, for example,
we assume that the each source contributes an equal number of plastic fragments ($N$),
then $A_k = b_k^{*3} N/\sigma = (c_k \Delta h_k)^3 N/\sigma $, and the resultant dependency of the inferred $\Delta h$ on $\Delta h_k$
is
\[
\Delta h =
\frac{b^*}{\rho A \ln(b^*/\Lambda)}
\sum_{k = 1}^K
\frac{\rho_k c_k^2 \Delta h_k^3 N}{\sigma}
   \ln\frac{c_k \Delta h_k}{\Lambda}.
\]
\color{black}
\clearpage

\section{\hlt{Three-dimensional model for fine microplastics}}\label{sec:3d_model}

\hlt{Our plate model presented in the Fracture model section and the Materials and Methods section (Fig.\,\ref{merged_fig1})
implicitly
allows for fragmented cells whose lateral size $\lambda$ is smaller than the thickness $\Delta h$ of the original plate.
It is not very realistic
to produce such small fragments from two dimensional fragmentation only
and we do not attempt to apply our two-dimensional fracture model directly to plastic fragments
for which $\lambda \ll \Delta h$.
We instead construct a three-dimensional version of our model to explain fine microplastics (${\sim}10\,\upmu$m$< \lambda < {\sim}300\,\upmu$m) recently observed in the upper ocean or on a beach\citep{EndersEA2015plastic,EoEA2018plastic,PoulainEA2018plastics,Pabortsava+Lampitt2020plastics}}
\begin{figure*}[h!]
  \centering\includegraphics[width=1.0\textwidth]{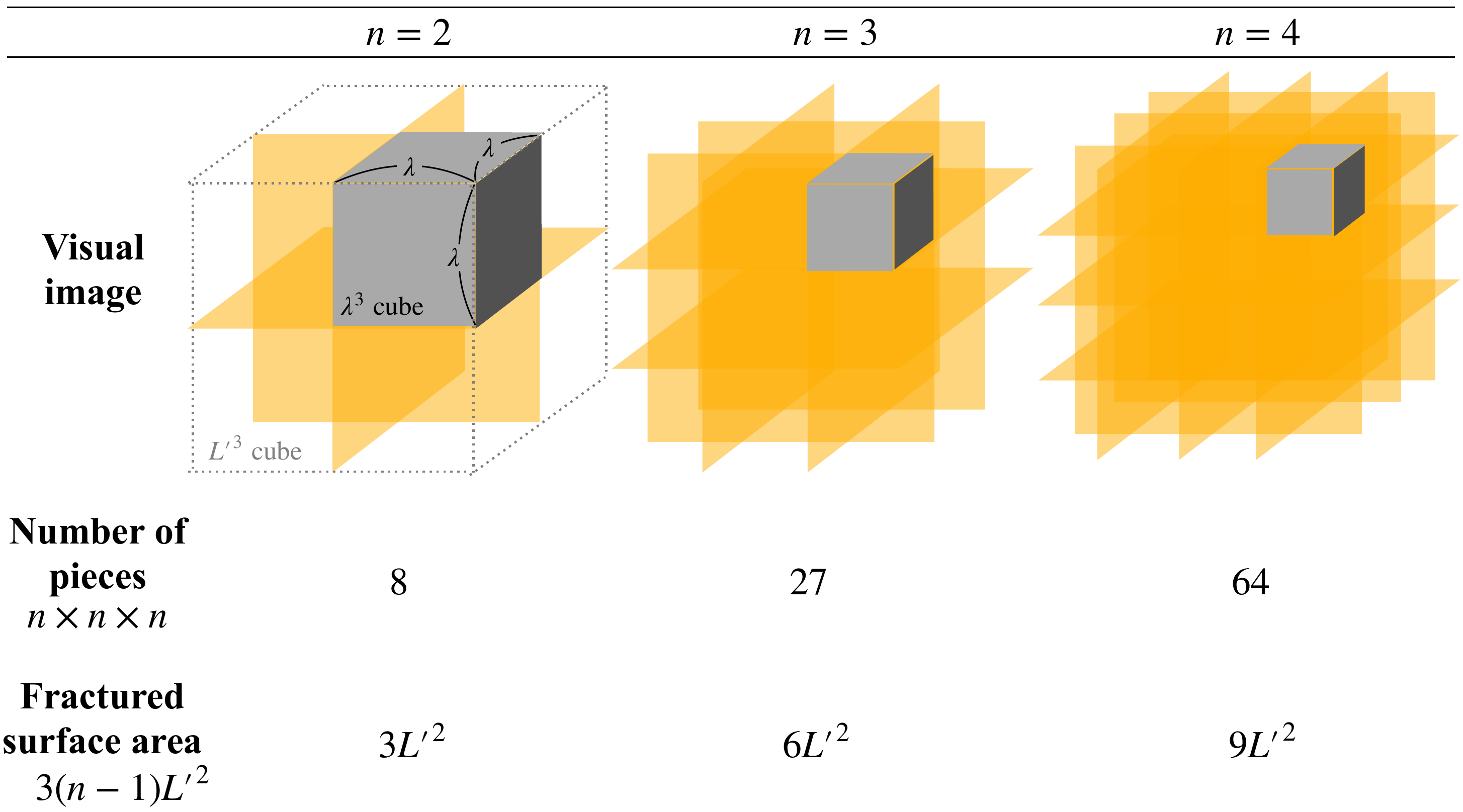}
\caption{\hlt{Schematic representation of 3-dimensional fracture model. All idealized plastic cube with a volume of $L'^3$ is broken into $n\times n\times n$ (middle) small cubes with the equal size of $\lambda=L'/n$. Orange planes show the surface created by the breakage and the total area of the planes depending on $n$ is shown at the bottom.}}
\label{fig_schem_3d-model}
\end{figure*}

\paragraph{\hlt{Three-dimensional model.}}

\hlt{As an extension
of the plate model,
we consider the fracture of a cube with a volume of $L'^3$ into $n\times n\times n$ equal-sized
cubic cells.
Shcematics are shown in Figure~\ref{fig_schem_3d-model}.
The cell size is then $\lambda=L'/n$;
with use of its inverse $\nu\equiv n/L'$, similarly to the original plate model, the number of pieces of the cells can be expressed as $L'^3\nu^3$.
Also, the area of
the new surfaces
produced in this breakage
is $3(n-1)L'^2$,
which is
proportional
to $n$ and hence
to $\nu$
when $n \gg 1$.
This allows defining the crush energy as $\varepsilon=jb\nu$, where $j$ is the number of $L'^3$ cubes to be fractured and $b\equiv3L'^3\phi$. Since this crush energy
is formally 
the same as for the original plate model \eqref{eq:crushenergy}, the expected number of
the fragments
is given by the same
Bose distribution  as~\eqref{eq:bose}. Thus, the size spectrum for the 3-dimensional model can be expressed as
\begin{align}
    \label{eq:size-spectrum-nu_3d}
    P(\nu)d\nu &= A\nu^3 \frac{1}{e^{\nu/\gamma^*}-1}d\nu
\quad\text{or} \\
    \label{eq:size-spectrum-lambda_3d}
    S(\lambda)d\lambda &= \frac{A}{\lambda^5} \frac{1}{e^{1/\lambda\gamma^*}-1}d\lambda,
\end{align}
where we have defined $\gamma^*\equiv \gamma/b$
as in the main manuscript.
These formulae are the same as \eqref{eq:size-spectrum-nu} and \eqref{eq:size-spectrum-lambda}
except for the exponents on
$\nu$ and $\lambda$, respectively. For a large size limit, i.e., $\lambda\gamma^*\gg1$, Eq.~\ref{eq:size-spectrum-lambda_3d} asymptotes to $A\gamma^*/\lambda^4$. Also, the peak size is approximately $\lambda_p \simeq 0.201 /\gamma^*$. In the similar fashion to the original plate model (See \ref{sec:total_abundance}), the total mass of the fragments in the size range from 0 to $\Lambda$ ($\Lambda \le L'$) is calculated as \begin{align}
    \label{eq:M-integral-3d}
    M &\equiv \int_{0}^{\Lambda} \rho\lambda^3 S(\lambda)d\lambda \nonumber\\
    &=-\rho A \gamma^* \ln (1-e^{-1/\gamma^*\Lambda}) \\
    &\simeq \rho A\gamma^* \ln (\gamma^*\Lambda).    
\end{align}
}

\paragraph{\hlt{Observed data.}}
\hlt{The 3-dimensional fracture model is applied to
four observed size distributions
of
fine microplastics smaller than $300\,\upmu$m.
First, we compare with the size distribution obtained in the North Atlantic Ocean by Pabortsava and Lampitt\citep{Pabortsava+Lampitt2020plastics} (PL2020). This data is the largest collection ($N=1444$) of
microplastics with sizes of $32\,\upmu$m\,$< \lambda < 651\,\upmu$m
in the wide depth range from 10\,m to 200\,m.
Since this collection, however, does not include
any data near the sea surface, we also compare with the observed size distributions in the North Atlantic Ocean obtained by Enders et al\citep{EndersEA2015plastic} (EN2015) and Poulain et al\citep{PoulainEA2018plastics} (PO2019) to complement this lack.
The former collected
finer microplastics with $11\,\upmu$m\,$<\lambda < 300\,\upmu$m sizes at $\sim$3\,m depth ($N=543$),
and the latter 
those with $25\,\upmu$m\,$< \lambda < 500\,\upmu$m
sizes in the water very close to the sea surface within 6\,cm ($N=520$).
Further, we compare with the observed size distribution ($20\,\upmu$m\,$< \lambda < 1000\,\upmu$m) obtained on Korean beach by Eo et al\cite{EoEA2018plastic} ($N=273738$), which could be considered as representative of the microplastics near the origin of their production. The size of the plastic pieces for all these observations is
identified by the longest dimension.
Additionally, EN2015 and PO2019 also use
the geometric mean $\sqrt{d_L d_W}$, where $d_L$ and $d_W$ denote the length and width.}

\paragraph{\hlt{Application.}}

\begin{figure*}[tp]
  \centering\includegraphics[width=1.0\textwidth]{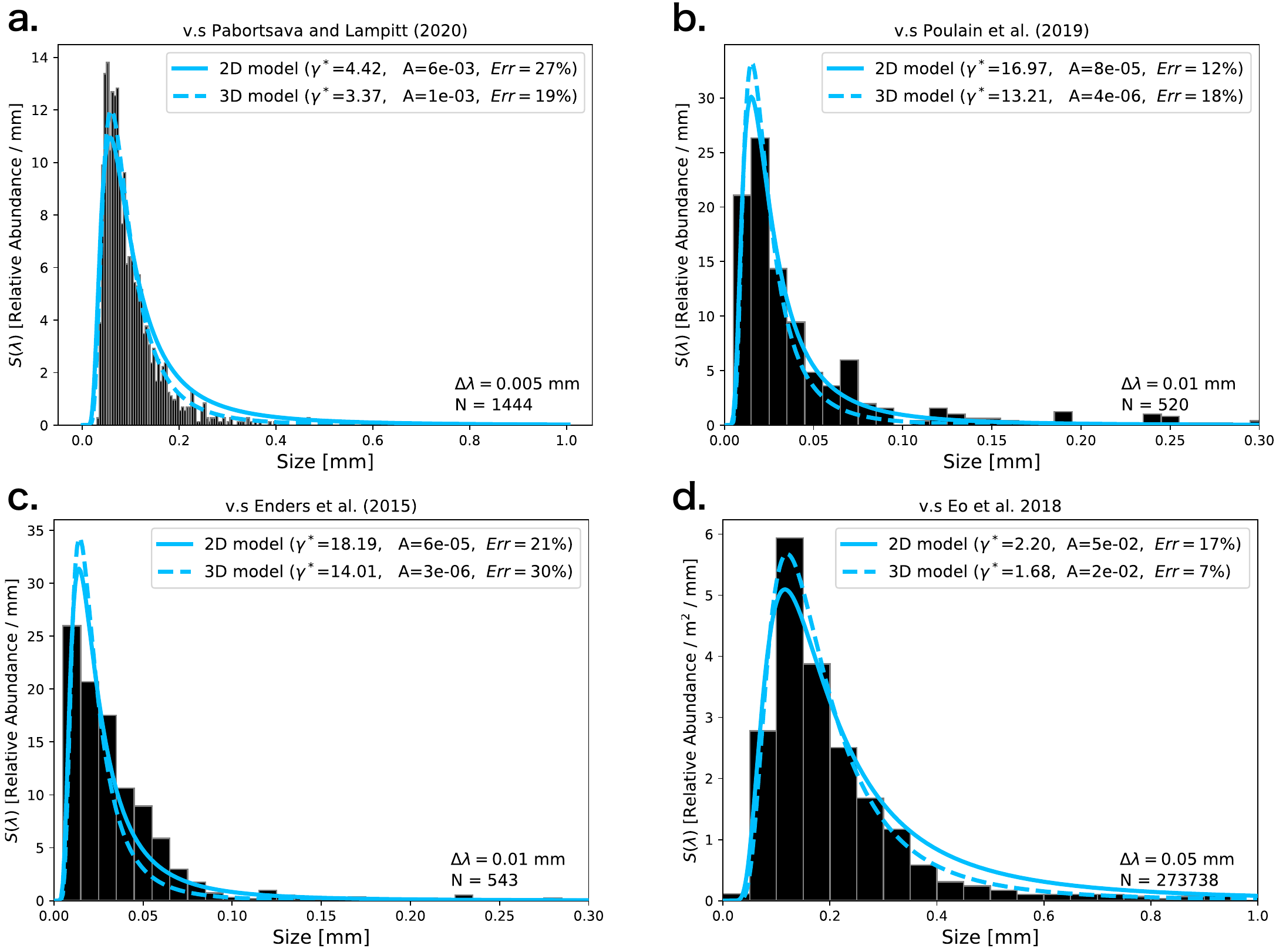}
\caption{\hlt{Size spectral densities of fine microplasitcs for observations (Black bars) and theories based on the original 2-dimensional plate model (Solid blue) and 3-dimensional model (Dashed blue). See Methods for conversion from histogram into size spectrum for the observed data. The observed data are obtained by digitizing the published original figures with WebPlotDigitizer (see the Materials and Methods section), except that the size distribution in {\bf a} is constructed from
the fragment size data. Pabortsava and Lampitt\citep{Pabortsava+Lampitt2020plastics} plot their data separately for a few depth ranges and polymer types. We have downloaded their data and merged all the data without any weights to construct the spectrum.
The spectrum is normalized by the total abundance N\@.
The sizes of the collected fragments for {\bf b} and {\bf c} are defined as geometric mean (see ``Observed data'' in this Appendix).}}
\label{fig_finemp1}
\end{figure*}

\hlt{The observed size distributions generally indicate an increase toward small size and sudden drop after passing the peak size  for all data except for that in EN2015 (Fig.~\ref{fig_finemp1}). This feature is qualitatively similar to that of the microplastics collected using the neuston net with the mesh size of 300\,$\upmu$m (cf.\ Figs.\,\ref{merged_fig2} and~\ref{merged_fig3}). Using the optimal $A$ and $\gamma^*$ obtained by the least square method over
the entire size range,
the theoretical curves of the 3-dimensional fracture model are found
to well match
the observed size distributions.
The optimal theoretical curve also reproduces the increase toward small size even for the size distribution in EN2015, which does not have a sudden drop. In this size range, the 3-dimensional fracture model seems to have higher reproducibility for PL2020 and EO2018 than
the 2-dimensional model
while the difference between the models is not clear for
the other cases.}

\begin{figure*}[t!]
  \centering\includegraphics[width=1.0\textwidth]{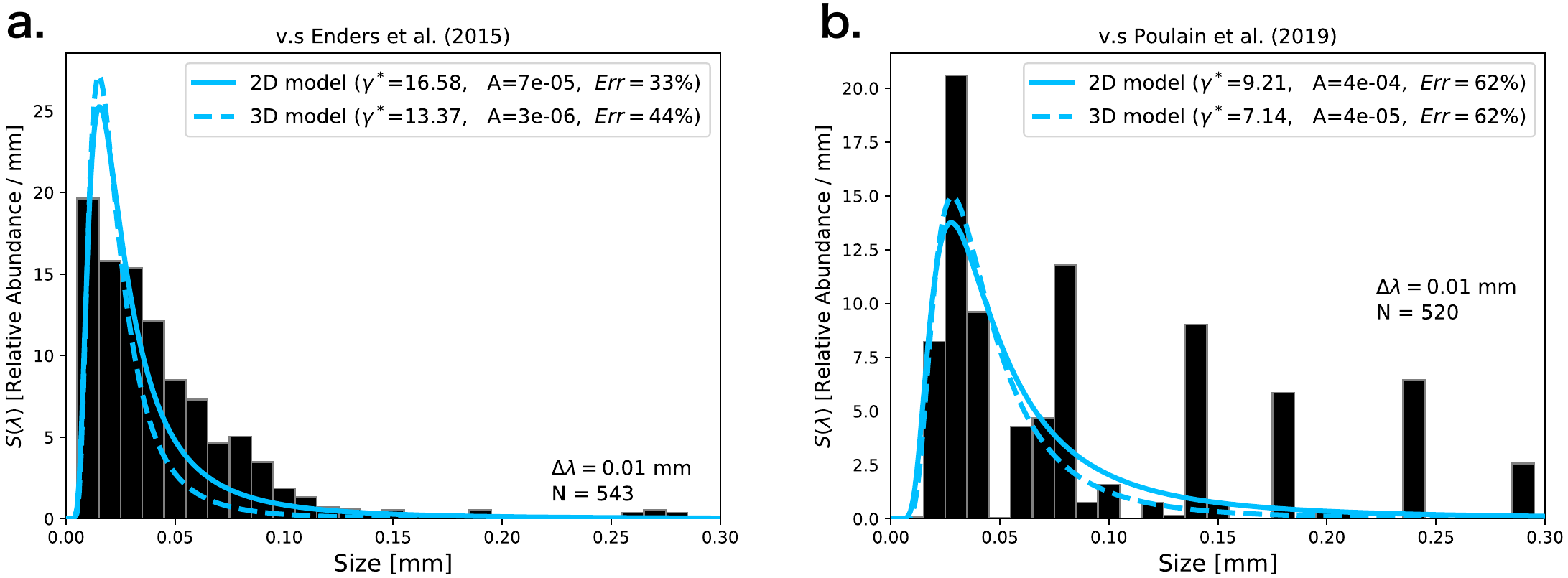}
\caption{\hlt{Panels~{\bf a} and~{\bf b} are respectively the same as Figs.\,\ref{fig_finemp1}c and \ref{fig_finemp1}b
except that the size of collected fragments
is defined as the longest dimension in these plots.}}
\label{fig_finemp2}
\end{figure*}

\hlt{Note that we have chosen geometric mean for size in Figs.\,\ref{fig_finemp1}c and~\ref{fig_finemp1}b.
When plotted with longest dimension, they are respectively Figs.\,\ref{fig_finemp2}a and~\ref{fig_finemp2}b.
Our model fits better with geometric mean (Figs.\,\ref{fig_finemp1}c and~\ref{fig_finemp1}b)
in both cases.
For EN2015, our model underestimates in the size range larger than 0.05\,mm (Fig.\,\ref{fig_finemp2}a)
and this tendency is weakened with geometric mean (Fig.\,\ref{fig_finemp1}c).
That is, geometric mean shifts larger fragments leftward in this case.
This is a natural consequence of geometric mean. 
For PO2019, geometric mean not only shifts the data leftward but also smooths the distribution.
The distribution is still not quite smooth, suggesting that more samples would be needed to get a smooth distribution.
Even with geometric mean, our model still underestimates for sizes larger than 0.05\,mm 
for EN2015 (Figs.\,\ref{fig_finemp1}c and \ref{fig_finemp2}a).
This might be because some of the samples were
collected using a 50\,$\upmu$m (0.05\,mm) mesh in some locations\citep{EndersEA2015plastic}.}

\clearpage
\section*{S1 Figure}\ \\
\setcounter{figure}{0}
\def\thefigure{S\arabic{figure} Figure}
\renewcommand{\figurename}{}

\begin{figure*}[h!]
  \centering\includegraphics[width=1.0\textwidth,clip]{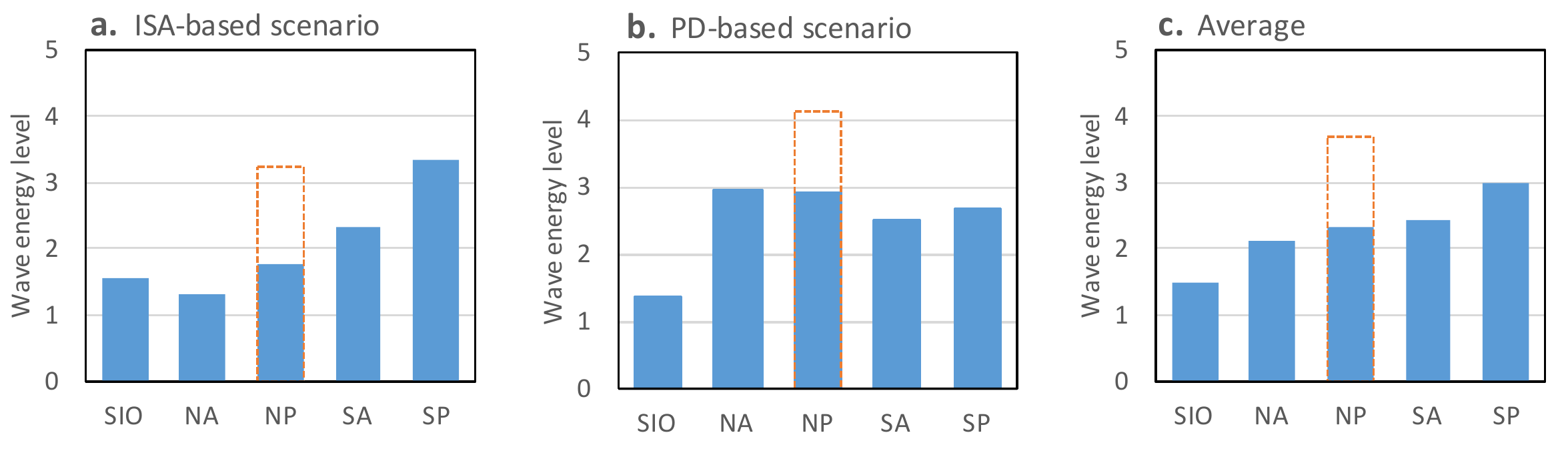}
\caption{Expected wave energy level \hlt{(no units)}
for accumulation zone in Southern Indian Ocean (SIO), North Atlantic (NA), North Pacific (NP), South Atlantic (SA), and South Pacific (SP) (blue bars). Dashed orange bars denote the case without the contribution from China in the North Pacific accumulation zone. }
\label{fig_meiji}
\end{figure*}
\clearpage

\FloatBarrier
\section*{S1 Table}\ \\
\setcounter{table}{0}
\def\thetable{S\arabic{table} Table}
\renewcommand{\tablename}{}

\begin{table}[h]
  \caption{Optimal $\gamma^*$ for different observation regions}
  \label{table_ex1}
  \centering
  \begin{tabular}{ccc}
    \hline
    {\bf Environmental energy} $\gamma^*$ [mm$^{-1}$] & {\bf Region}  & {\bf Literature}  \\
    \hline
    0.24  & {\small North Atlantic Ocean}   &  {\small C{\'o}zar et al. 2014}\\
    \hline
    0.24 & {\small Around Japan} & {\small Isobe et al. 2015}\\
    \hline
    0.26 & {\small Western Pacific transoceanic section} & {\small Isobe et al. 2019}\\
    \hline
    0.27 & {\small South Indian Ocean} & {\small C{\'o}zar et al. 2014} \\
    \hline
    0.27  & {\small South Atlantic Ocean}  &  {\small C{\'o}zar et al. 2014}\\
    \hline
    0.35 & {\small North Pacific Ocean} & {\small C{\'o}zar et al. 2014}\\
    \hline
    0.35 & {\small South Pacific Ocean} & {\small C{\'o}zar et al. 2014}\\
    \hline
    0.39 & {\small Seto Inland Sea} & {\small Isobe et al. 2014}\\
    \hline
  \end{tabular}
\end{table}
\clearpage